\PassOptionsToPackage{unicode}{hyperref}
\PassOptionsToPackage{hyphens}{url}
\PassOptionsToPackage{dvipsnames,svgnames,x11names}{xcolor}
\documentclass[
  12pt]{article}

\usepackage{amsmath,amssymb}
\usepackage{iftex}
\ifPDFTeX
  \usepackage[T1]{fontenc}
  \usepackage[utf8]{inputenc}
  \usepackage{textcomp} 
\else 
  \usepackage{unicode-math}
  \defaultfontfeatures{Scale=MatchLowercase}
  \defaultfontfeatures[\rmfamily]{Ligatures=TeX,Scale=1}
\fi
\usepackage{lmodern}
\ifPDFTeX\else
\fi
\IfFileExists{upquote.sty}{\usepackage{upquote}}{}
\IfFileExists{microtype.sty}{
  \usepackage[]{microtype}
  \UseMicrotypeSet[protrusion]{basicmath} 
}{}
\makeatletter
\@ifundefined{KOMAClassName}{
  \IfFileExists{parskip.sty}{%
    \usepackage{parskip}
  }{
    \setlength{\parindent}{0pt}
    \setlength{\parskip}{6pt plus 2pt minus 1pt}}
}{
  \KOMAoptions{parskip=half}}
\makeatother
\usepackage{xcolor}
\makeatletter
\ifx\paragraph\undefined\else
  \let\oldparagraph\paragraph
  \renewcommand{\paragraph}{
    \@ifstar
      \xxxParagraphStar
      \xxxParagraphNoStar
  }
  \newcommand{\xxxParagraphStar}[1]{\oldparagraph*{#1}\mbox{}}
  \newcommand{\xxxParagraphNoStar}[1]{\oldparagraph{#1}\mbox{}}
\fi
\ifx\subparagraph\undefined\else
  \let\oldsubparagraph\subparagraph
  \renewcommand{\subparagraph}{
    \@ifstar
      \xxxSubParagraphStar
      \xxxSubParagraphNoStar
  }
  \newcommand{\xxxSubParagraphStar}[1]{\oldsubparagraph*{#1}\mbox{}}
  \newcommand{\xxxSubParagraphNoStar}[1]{\oldsubparagraph{#1}\mbox{}}
\fi
\makeatother

\usepackage{longtable,booktabs,array}
\usepackage{calc} 
\usepackage{etoolbox}
\makeatletter
\patchcmd\longtable{\par}{\if@noskipsec\mbox{}\fi\par}{}{}
\makeatother
\IfFileExists{footnotehyper.sty}{\usepackage{footnotehyper}}{\usepackage{footnote}}
\makesavenoteenv{longtable}
\usepackage{graphicx}
\makeatletter
\def\maxwidth{\ifdim\Gin@nat@width>\linewidth\linewidth\else\Gin@nat@width\fi}
\def\maxheight{\ifdim\Gin@nat@height>\textheight\textheight\else\Gin@nat@height\fi}
\makeatother
\setkeys{Gin}{width=\maxwidth,height=\maxheight,keepaspectratio}
\makeatletter
\def\fps@figure{htbp}
\makeatother

\makeatletter
\@ifpackageloaded{caption}{}{\usepackage{caption}}
\AtBeginDocument{%
\ifdefined\contentsname
  \renewcommand*\contentsname{Table of contents}
\else
  \newcommand\contentsname{Table of contents}
\fi
\ifdefined\listfigurename
  \renewcommand*\listfigurename{List of Figures}
\else
  \newcommand\listfigurename{List of Figures}
\fi
\ifdefined\listtablename
  \renewcommand*\listtablename{List of Tables}
\else
  \newcommand\listtablename{List of Tables}
\fi
\ifdefined\figurename
  \renewcommand*\figurename{Figure}
\else
  \newcommand\figurename{Figure}
\fi
\ifdefined\tablename
  \renewcommand*\tablename{Table}
\else
  \newcommand\tablename{Table}
\fi
}
\@ifpackageloaded{float}{}{\usepackage{float}}
\floatstyle{ruled}
\@ifundefined{c@chapter}{\newfloat{codelisting}{h}{lop}}{\newfloat{codelisting}{h}{lop}[chapter]}
\floatname{codelisting}{Listing}

\makeatother
\makeatletter
\@ifpackageloaded{caption}{}{\usepackage{caption}}
\@ifpackageloaded{subcaption}{}{\usepackage{subcaption}}
\makeatother

\ifLuaTeX
  \usepackage{selnolig}  
\fi
\usepackage[]{natbib}
\usepackage{bookmark}

\IfFileExists{xurl.sty}{\usepackage{xurl}}{} 
\hypersetup{
  pdftitle={Direction-Preferring Confidence Intervals},
  pdfauthor={},
  pdfkeywords={selective inference, false coverage-statement rate, sign determination, confidence-interval length},
  colorlinks=true,
  linkcolor={blue},
  filecolor={Maroon},
  citecolor={Blue},
  urlcolor={Blue},
  pdfcreator={LaTeX via pandoc}}

\newcommand{\anon}{0}

\usepackage{amsthm}
\usepackage{graphicx,psfrag,epsf}
\usepackage{enumerate}
\usepackage{natbib}
\usepackage{svg}
\usepackage{url} 
\usepackage{multirow}
\usepackage{authblk}
\usepackage{hyperref}
\usepackage{cleveref}
\usepackage{algorithm}
\usepackage{algpseudocode}
\newcommand{\blind}{1}

\newcommand{\Ic}{\mathcal{I}^c}
\newcommand{\Imar}{\mathcal{I}}
\newcommand{\Ac}{\mathcal{A}^c}
\newcommand{\Am}{\mathcal{A}}

\newcommand{\Fc}{F^c_\theta}
\newcommand{\ns}{std}
\newcommand{\DPp}{DP\ensuremath{^+} }
\newcommand{\mI}{\ensuremath{\mathbb I}}
\newcommand{\mE}{\ensuremath{\mathbb E}}
\newcommand{\mP}{\ensuremath{\mathbb P}}

\newcommand{\rev}[1]{{\color{black}#1}}

\newtheorem{theorem}{Theorem}[section]

\newtheorem{corollary}[theorem]{Corollary}
\newtheorem{proposition}[theorem]{Proposition}

\newtheorem{remark}[theorem]{Remark}

\begin{document}

\def\spacingset#1{\renewcommand{\baselinestretch}%
{#1}\small\normalsize} \spacingset{1}



\if0\anon
{
 \title{\bf Direction-Preferring Confidence Intervals}
 \author[1]{ Tzviel Frostig}
 \author[2]{Orestis A. Panagiotou}
 \author[1]{ Yoav Benjamini}
 \author[1]{ Ruth Heller}
 \affil[1]{ Department of Statistics and Operations Research, Tel Aviv University, Tel Aviv, Israel}
 \affil[2]{Department of Health Services, Policy \& Practice, Brown University School of Public Health, Providence, USA}

 \maketitle
} \fi

\if1\blind
{
 \bigskip
 \bigskip
 \bigskip
 \begin{center}
  {\LARGE\bf Direction-Preferring Confidence Intervals}
\end{center}
 \medskip
} \fi

\bigskip


\begin{abstract}
Confidence intervals (CIs) are fundamental in statistical analysis, providing interval estimates for unknown parameters. We introduce direction-preferring CIs that prioritize inference for parameters with a specific sign (e.g., positive sign). Unlike previous methods of constructing CIs, our intervals are based on the inversion of acceptance regions that differ for positive and negative parameters, to enhance the probability of determining the sign in the prespecified preferred direction while
maintaining standard coverage. 
We apply our suggestion in the typical context that multiple parameters are considered, but  only some of the constructed CIs  are of interest, e.g.,
 when CIs are reported only for the subset of parameters that are found to differ significantly from zero.
 This leads us to suggest direction-preferring CIs with false coverage rate (FCR) control, as well as conditional direction-preferring CIs that adjust for selection by conditioning on the selection event. 
We compare the two selection adjustment methods in theory and using numerical experiments. The procedures we develop are useful for inference on secondary outcomes in medical applications, a setting we highlight through two motivating examples. 
\end{abstract}

\noindent%
\vfill

\newpage
\spacingset{1.45} 

\section{Introduction}


Confidence intervals (CIs) have gained renewed prominence across scientific applications, sometimes replacing p-values or formal hypothesis testing. In such cases CIs are expected to serve as the sole inferential summary  of the results  of clinical trials or observational studies. \cite{tukey1991philosophy} lists three questions, in order of importance, to be answered sequentially about a parameter within the limit of acceptable uncertainty. i) What is the direction of the effect? ii) What is the minimum effect size? iii) What is the maximum effect size? He then turns to the use of CIs to answer these questions. If the sign of the effect (positive or negative) is indeterminable, a short two-sided interval is preferable. Conversely, if the sign is clear (suppose that it is positive), the lower end of the CI helps assess the minimal effect, ideally being as far from zero as possible. The upper end of the interval, though less critical, estimates the maximal effect and should certainly be finite and not too far away. The main thrust of that work by Tukey was to design CIs when multiple parameters are considered. Addressing multiplicity requires the inflation of the CIs' lengths which come at the expense of sign-determination. While Tukey's formulation of the three goals has appealed to many, the compromises he designed have not received much attention to date. How to compromise between sign determination, minimal effect size and length, while facing multiplicity, is the subject of this work.

The standard method for constructing a (single) CI for a location family is to invert acceptance regions for testing the parameter. The acceptance regions are equivariant if their shape does not depend on the value of the parameter tested. Inverting equivariant acceptance regions results in a CI that is also equivariant, having the same shape around the estimate for any value of the estimate, and thus a fixed length regardless of the potential for sign determination.
Non-equivariant acceptance regions allow the designer to weight differently Tukey's goals, extending the length of the CI in exchange for better sign determination or vice versa \citep{pratt1961length, hayter1994relationship, benjamini1996nonequivariant}.

The non-equivariant methods used for building CIs typically treat the two  directions symmetrically, by reflection: if two estimators turned out to be the same with opposite signs, one CI was the exact reflection of the other. For an asymmetric treatment, a simple method is to use asymmetric equivariant CIs that allocate  $\alpha$ unequally for the lower and upper bounds.
We consider here the advantage offered by constructing acceptance regions in the preferred direction that are shorter than the acceptance regions in the non-preferred direction. Compared with asymmetric  equivariant CIs, the direction-preferring CIs are substantially shorter (see details in \S~\ref{sect:comp_equivariant}).  In addition, we examine the performance of these direction-preferring CIs when combined with adjustment for multiple inferences.

For constructing CIs for the selected parameters from the larger family of hypotheses, rather than using the familywise error rate (FWER)
, we can use the concept of the false coverage rate (FCR) of \cite{benjamini2005false}, which is the expected proportion of non-covering intervals over the selected.
Their suggested way of constructing CIs can be used with any selection method, as long as the method constructed at level $\alpha$ has coverage at least $1-\alpha$, and is monotone in $\alpha$ (i.e., for all $0<\alpha<\alpha'<1$, the CI at level $\alpha$ contains that at level $\alpha'$, see \citealt{weinstein2020selective} for details and examples).

Another way is to condition on the parameter being selected according to a specified selection criterion, and report the CI conditional on being selected. The procedure that reports the conditional CIs selected also controls the FCR, see \cite{weinstein2013selection}.

We will apply these two methods in conjunction with the new direction-preferring intervals. In particular, the CIs proposed here aim to balance the conflicting goals inherent in the analysis of secondary outcomes in medical research, discussed in \S~\ref{subsec:intro-motivation}.

Our main contributions are the following:
\begin{itemize}
  \item \textbf{Marginal Direction-Preferring CIs.}
We propose CIs with a preferred direction (e.g., positive) that offer improved sign-determination power, shorter lengths, and more accurate effect size estimation in the preferred direction, while still allowing detection of non-preferred (e.g., negative) signals. This is achieved by inverting acceptance regions that are longer for the non-preferred direction than for the preferred direction.
 \item \textbf{Conditional Direction-Preferring CIs.}
 We propose conditional CIs with a preferred direction that share the same building blocks as the marginal CIs. These can be used for secondary or exploratory outcomes that were not pre-registered.
\item \textbf{Multiple Direction-Preferring CIs.}
We address the problem of constructing direction-preferring CIs for selected parameters. We suggest methods for adjusting for selection by reducing the error level in which the marginal direction-preferring CIs are constructed so the FCR is controlled or through conditional coverage.
\item \textbf{Comparing multiple Direction-Preferring CIs.}
We show that conditional CIs will have on average less power to determine the sign compared to the approach in \cite{benjamini2005false}. However, their theoretical guarantee is stronger, and their construction is unaffected by the number of parameters initially considered, unlike the approach in \cite{benjamini2005false}.
\item \textbf{Clinical case studies.} We demonstrate the utility of the methods for secondary outcomes and apply them to the VITAL and SURPASS-CVOT trials.
\end{itemize}

\subsection{Motivating application: Secondary outcomes in clinical trials}\label{subsec:intro-motivation}

The development of our CIs is motivated by the newly emphasized role of CIs in medical applications. Sometimes, CIs are expected to answer solely the challenges of reporting the results of clinical trials or observational studies. The interest primarily lies in whether the treatment is beneficial, and then in quantifying its minimum and maximum level of efficacy, while it is still of value to offer some quantitative assessment if the treatment is not beneficial. 
It is also rarely the case that only one measure of efficacy is considered in medical research, raising the second challenge of multiplicity. As a first step to address the challenge, these multiple outcomes are typically classified into very few primary outcomes and numerous secondary outcomes.

The "New guidelines for reporting results" in the New England Journal of Medicine (NEJM) and their discussion starting with the editorial \citep{harrington2019new} offer an important example of the trend we described.
For the primary outcomes the requirement is to report two-sided CIs to convey effect sizes and the uncertainty involved, and to report p-values and adjust for multiplicity when performing formal tests.
For the secondary outcomes, if no protocol or prespecified analysis plan exists for addressing the multiplicity issue, the requirement is to report only two-sided 95\% CIs, with no p-values or formal testing \citep{harrington2019new}.

As an illustration of their recommendation to report only nominal 95\% CIs, \cite{harrington2019new} discuss the VITAL trial \citep{Manson2019MarineN3VITAL}, a large randomized study of the potential health benefits of fish-oil (marine $n-3$) supplementation.
For the two prespecified primary outcomes, both p-values and marginal 95\% CIs were reported. There were 22 secondary outcomes as counted by \cite{harrington2019new} for which only 95\% CIs were reported. Even though some p-values of secondary outcomes were below 0.05, they were not reported, per the new guidelines. We shall use this example throughout the paper to demonstrate and compare the different CIs proposed and discussed.

Revisiting \cite{Manson2019MarineN3VITAL}, we noticed that the authors admit in the summary that the results do not show the usefulness of fish-oil supplement.
Still, they highlighted in their discussion one secondary outcome, myocardial infarction, for which the 95\% CI excluded the value of no effect, stating that the risk ``was lower in the $n-3$ group than in the placebo group". The results of this trial were later discussed in a meta-analysis. \cite{Wu2019DietaryFatsCardiometabolic} subsequently cited  the four secondary outcomes whose $95\%$ CIs excluded the no-effect value as evidence that $n-3$ fatty acid supplementation reduces risk. These statements ignore the data-driven selection of outcomes.

This illustrates the problem in the new guidelines regarding the reporting of secondary outcomes: in practice, analysts typically focus on the subset of secondary outcomes whose CIs exclude the value of no-effect, even if not in a formal way. This data‐driven selection undermines the nominal guarantee. Although 95\% CIs ensure an average non-coverage rate of 5\% across all outcomes, the average non-coverage rate across the selected outcomes that, i.e., those CIs that exclude the null value, can be substantially higher.  So the effect of selection cannot be ignored when reporting CIs.

Responding to the guidelines, \rev{\cite{betensky}} argued that the absence of a prespecified multiple testing procedure need not preclude a multiplicity-adjusted analysis,  by demonstrating  that almost all multiple testing methods control the increased error from multiple testing if the treatment has no advantage at all over the control. 
In his answer, \cite{harrington2019reply} defended the dropping of $p$-values and multiplicity adjusted testing, arguing that the control of FWER over all endpoints will reduce the power of the primary test to discover the treatment effect, and that the family of secondary outcomes may include some which were not defined prior to the study, hindering the use of the current multiplicity adjustments that require a prespecified family of endpoints.

What comes out of the above demonstration and discussion of the analysis of secondary outcomes is that:

(i) The analysts and their readers are interested in the two-sided CIs for the most interesting results, which correspond to the estimates that avoid the no-effect value;

(ii) Addressing multiplicity is considered important in the face of such selection,
but possibly not as much as it is for the primary endpoints;

(iii) It is important to be able to keep power to detect an effect in the direction of improvement.

We thus see that these are essentially the issues that occupied \cite{tukey1991philosophy} as discussed in the opening section, with the added emphasis on a prespecified direction. Added to these is,

(iv) 
Some scientifically relevant outcomes may not belong to a prespecified multiplicity family, motivating a conditional approach when the target and selection event are clearly defined.

Addressing the concern about selective inference, \cite{benjamini1998confidence} used the simultaneous approach to offer better separation from zero while ensuring simultaneous coverage.
Still, the improvements in sign determination and separation from zero were mainly useful for a small number of parameters, so these may be more relevant to the analysis of primary endpoints. Importantly, \cite{benjamini1998confidence} did not address the setting where one direction is preferred over  the other.


The rest of the manuscript is organized as follows. In \S~\ref{sec-generalized-marginal} and \S~\ref{sec-conditional}, we present the marginal and conditional direction-preferring CIs, provide a qualitative comparison with other relevant methods for constructing CIs, and illustrate their application to the 22 secondary outcomes in the study of \cite{Manson2019MarineN3VITAL}.
 In \S~\ref{sec-single-comparison} we provide  detailed numerical experiments. In \S~\ref{sect:selected_paramters} we construct direction-preferring CIs following selection, and compare the conditional and non-conditional approach theoretically as well as numerically. In \S~\ref{sect:real_example} we apply our suggested methodology to the results of the SURPASS-CVOT trial \citep{Nicholls2025Cardiovascular} for additional illustration, and in \S~\ref{sect:disc} we conclude with final remarks.

\section{Generalized Marginal Confidence Intervals}\label{sec-generalized-marginal}


Let $Y$ be our point estimate for the location parameter $\theta$. Throughout the paper we assume that $Y$ follows a location family distribution, i.e., $F_\theta(y) = F(y - \theta)$, where $\theta \in \mathbb{R}$ and $F_\theta$ is the cumulative distribution function of $Y$. For simplicity we assume that $F_0$ is continuous and strictly increasing and has an absolutely continuous density $f_0$, and that $f_0$ is unimodal and symmetric. With slight adjustments, the CIs presented can be modified to accommodate other types of distributions as well. Let $f_\theta(y)$ and
$q_\theta(1 - \alpha)$ be, respectively, the density and the $1-\alpha$ quantile of $Y$, where $q_\theta(p) = \left( F_{\theta}\right)^{-1}(p)$. To simplify notation, we denote by $q_{p} = q_0(p)$.

We consider methods to construct a $1-\alpha$ CI for $\theta$ by inverting acceptance regions of level $\alpha$ hypotheses tests. Let $\Am(\tau; \alpha)$ be the acceptance region for testing $H_0: \theta = \tau$ at significance level $\alpha$. The confidence region constructed by inverting the acceptance region \citep{lehmann2005testing} at confidence level $1 - \alpha$ is $$ \Am^{-1}(Y; \alpha) = \{ \tau : Y \in \Am(\tau; \alpha)\}. $$
When $\Am^{-1}$ is disconnected, the CIs are taken as the convex hull of $\Am^{-1}$, denoted by $\Imar (Y;\alpha)$, or simply $\Imar$ when the meaning is clear. 


In the application to secondary outcomes we rely on the normal approximation for $Y_i = \hat \beta_i/se_i,$ where $\hat \beta_i$ is the effect estimate for outcome $i$ and $se_i$ is its standard error or an estimate of the standard error.

\subsection{Review of standard and modified Pratt CIs}\label{subsec-review-marginal}
\cite{pratt1961length} showed the following relation between  the expected total length of a $1-\alpha$ confidence region constructed by inverting acceptance regions of level $\alpha$ tests: 
\begin{equation}
\mE_{\theta}\left[\lambda\{\mathcal A^{-1}(Y; \alpha) \} \right] = \int_{\tau\neq \theta}\mP_{\theta}(Y\in \mathcal A(\tau; \alpha))d\tau, \label{eq-pratt-identity}
\end{equation}
where $\lambda(B)$ denotes the Lebesgue measure of a set $B\subseteq\mathbb R$.
Pratt's identity shows that expected total length at a fixed true value $\theta$ is reduced by  constructing acceptance regions that provide the most powerful test at $\tau\neq \theta$, but these regions vary with the null parameter value $\tau$. For example, for $\theta=0$, the most powerful test of the null hypothesis that the parameter value is $\tau$ is one-sided, but it points in opposite directions for  $\tau>0$ and for $\tau<0$.

A natural approach to consider is the shortest  region,  defined as $\Am_{\ns}(\theta) = \{y: f_\theta(y) \geq \xi \}$, where $\xi$ is the maximum value ensuring that $ \mathbb{P}_{\theta}(Y \in \Am_{\ns}(\theta)) = 1 - \alpha$.
For a symmetric unimodal distribution, it is represented as$$ \Am_{\ns}(\theta ; \alpha) = (\theta-q_{1 - \alpha / 2}, \theta+ q_{1-\alpha / 2}).$$ 
Inverting the acceptance region results in the standard shortest CI, \begin{equation} \label{eq:ci_standard}
\Imar_{\ns}(Y;\alpha) = (Y - q_{1-\alpha / 2}, Y + q_{1 - \alpha / 2}),
\end{equation}



Interested in a CI of minimal length when the parameter is at 0,  \cite{pratt1961length} further used the identity in \eqref{eq-pratt-identity} for this purpose. These CIs turn out to be equivalent to Neyman's three-decision rule that determine the sign when $|Y| > q_{1-\alpha}$. Alas, these CIs grow in length as they get further from zero, with no limit. \cite{benjamini1998confidence} pointed out that this property is unattractive, as did Pratt himself, and they suggested instead the modified Pratt (MP) CIs, which provide better directional conclusions than standard CIs and have bounded lengths.
The idea behind the MP CI is to allow the inflation of up to $r$ in the acceptance region length compared to the standard acceptance region. Specifically, compared to the shortest acceptance region: for $\theta>0$, the lower bound of the acceptance region is higher; for $\theta<0$, the upper bound of the acceptance region is lower (see Figure \ref{fig:all_ci_ar}).
 For completeness, a formal description of the CI is provided in \S~ \ref{sect:description_full}
 .

\subsection{Direction-Preferring Confidence Intervals} \label{sect:direction_ci}

The CIs discussed in \S~\ref{subsec-review-marginal} treat positive and negative parameters symmetrically. We now introduce direction-preferring CIs that reflect  a predisposition toward the sign of the parameter. When describing direction-preferring CIs in the positive direction, we refer to the resulting construction as $DP^+$. The construction preferring the negative direction follows analogously by reversing the roles of positive and negative $\theta$, and is denoted by  $DP^-$.

For the positive-preferring construction: when $\theta > 0$, the analyst seeks higher power to detect that the parameter value is positive and a CI that is as short as possible;  when $\theta \leq 0$, by contrast, the analyst allows for a longer CI, provided it reliably excludes positive values (i.e., establishes $\theta \leq 0$).



Since we invert acceptance regions to construct the CI, the interval will be entirely above zero only if the observed value $Y$ is above the largest upper bound on the acceptance region for non-positive parameters.
For best positive-sign determination, we should thus modify the acceptance region when testing that the value is $\tau\leq 0$ to be one sided (so that the upper bound of the acceptance region is the lowest possible, while still guaranteeing that the probability of being in the acceptance region is $1-\alpha$). The acceptance region is therefore:
\[
\mathcal A_{\DPp}(\tau; \alpha) =
\begin{cases}
\mathcal A_{\ns}(\tau; \alpha) & \text{if } \tau>0,\\[6pt]
\bigl(-\infty,\tau+q_{1-\alpha} \bigr) & \text{if } \tau\leq 0.
\end{cases}
\]

The resulting CI is:
\[
\mathcal I(Y; \alpha)=
\begin{cases}
\bigl[ Y - q_{1-\alpha/2},\; Y + q_{1-\alpha/2} \bigr], & \text{if } Y > q_{1-\alpha/2},\\[6pt]
\bigl( 0,\; Y + q_{1-\alpha/2} \bigr], & \text{if } q_{1-\alpha} < Y \le q_{1-\alpha/2},\\[6pt]
\bigl[ Y - q_{1-\alpha},\; \max\{0,\; Y + q_{1-\alpha/2}\} \bigr], & \text{if } Y \le q_{1-\alpha}.
\end{cases}
\]

Compared with $\mathcal I_{std}(Y; \alpha)$, the main advantage is that we obtain a shorter CI while still achieving positive sign determination for $q_{1-\alpha} < Y \leq q_{1-\alpha/2}$. The trade-off is that for $Y \leq q_{1-\alpha}$ the CI becomes longer, with its length increasing as $Y$ decreases. To address this limitation, following \cite{benjamini1998confidence}, we constrain the length of the acceptance region in the nonpreferred direction relative to the standard acceptance region  \rev{using the inflation factor $r$ and its derived tail allocation $\epsilon_{marg}(r,\alpha)$ defined below}.
Furthermore, for sufficiently negative parameter values, no inflation is applied to the acceptance regions. Consequently, for sufficiently small values of $Y$, the resulting CI aligns with the standard CI. 

\rev{We parameterize the construction by the acceptance-region inflation factor $r\geq1$. Let $e=\epsilon_{marg}(r,\alpha)\in[0,\alpha/2]$ be the unique solution of
\[
q_{1-\alpha+e}-q_{e}=2r q_{1-\alpha/2}.
\]
The direction-preferring acceptance region (in the positive direction), $\Am_{\DPp}$, is then as follows:}
\begin{equation} \label{eq:acceptance_amp}
{\mathcal A_{\DPp}(\theta;r,\alpha)=
\begin{cases}
\bigl( \theta - q_{1-\alpha/2},\; \theta + q_{1-\alpha/2} \bigr) & \text{if } \theta > 0,\\[6pt]
\bigl( \theta - q_{1-e},\; \theta + q_{1-\alpha+e} \bigr) & \text{if } -q_{1-\alpha/2} < \theta \le 0,\\[6pt]
\bigl( \theta - q_{1-\alpha/2},\; \theta + q_{1-\alpha/2} \bigr) & \text{if } \theta \le -q_{1-\alpha/2}.
\end{cases}}
\end{equation}

Note that the acceptance region for $\theta \leq -q_{1-\alpha/2}$ is the same as the standard acceptance region, rather than the longer acceptance region for $-q_{1-\alpha/2}<\theta\leq 0$, since for $\theta \leq -q_{1-\alpha/2}$  the upper bound is nonpositive, so it has no impact on the lower bound of the CI for a positive test statistic $Y>0$.

\rev{The resulting CI is:}
\begin{equation} \label{eq:pp-marginal}
{\mathcal I_{\DPp,r}(Y;\alpha) \;=\;
\begin{cases}
\bigl[\,Y - q_{1-\alpha/2},\;\; Y + q_{1-\alpha/2}\,\bigr]
& \text{if } Y \ge q_{1-\alpha/2}, \\[8pt]
\bigl(\,0,\;\; Y + q_{1-\alpha/2}\,\bigr]
& \text{if } q_{1-\alpha+e} \le Y < q_{1-\alpha/2}, \\[8pt]
\bigl[\,Y - q_{1-\alpha+e},\;\; Y + q_{1-\alpha/2}\,\bigr]
& \text{if } 0 \le Y < q_{1-\alpha+e}, \\[8pt]
\bigl[\,Y - q_{1-\alpha/2},\;\; Y + q_{1-\alpha/2}\,\bigr]
& \text{if } -q_{1-\alpha/2} \le Y < 0,\\[8pt]
\bigl[\,Y - q_{1-\alpha/2},\;\; 0\,\bigr]
& \text{if } -q_{1-e}\le Y< -q_{1-\alpha/2}, \\[8pt]
\bigl[\,Y - q_{1-\alpha/2},\;\; Y+q_{1-e}\,\bigr]
& \text{if } -q_{1-\alpha/2}-q_{1-e}\le Y< -q_{1-e}, \\[8pt]
\bigl[\,Y - q_{1-\alpha/2},\;\; Y + q_{1-\alpha/2}\,\bigr]
& \text{if } Y<-q_{1-\alpha/2}-q_{1-e} .
\end{cases}}
\end{equation}

\begin{figure}
 \centering
  \centering
 \includegraphics[width=0.7\textwidth,height=5cm]{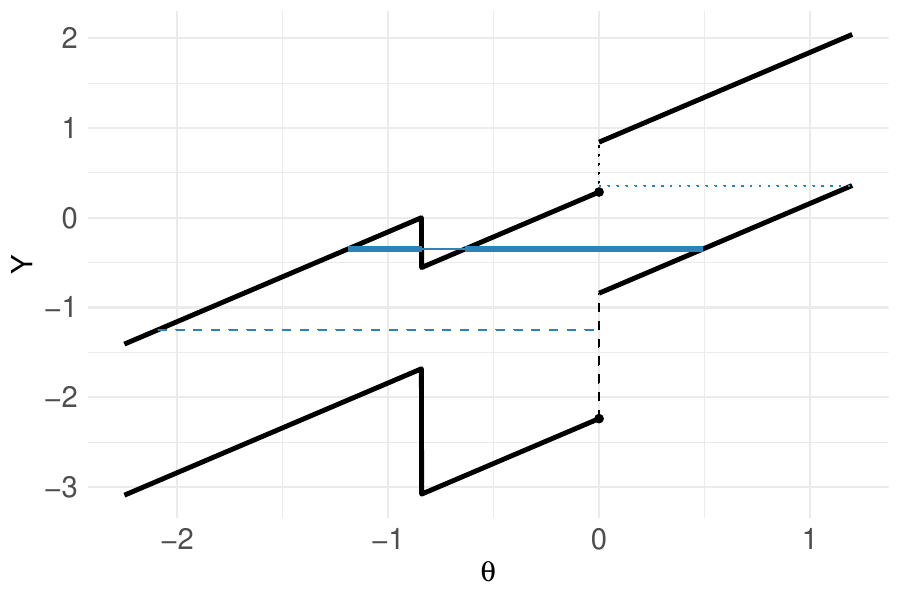}
 \caption{\small For a test statistic $Y\sim N(\theta,1)$, we illustrate positive-preferring acceptance regions given by the bold lines and some resulting CIs. The level $\alpha=0.4$ is chosen for readability.
 \rev{The parameter governing the preference is $r=1.50$, for which $\epsilon_{marg}(r,\alpha)=0.0126$.} The acceptance region for $\theta=0$ ranges from $-2.23$ to $0.28$, extending towards negative values. 
 At $\theta = -0.84$ the acceptance region reverts to its shortest. The discontinuity at $\theta = 0$ causes the acceptance region upper bound to jump from $0.28$ to $0.84$ when $\theta$ increases slightly.
 The blue horizontal lines represent CIs from inverted acceptance regions: dotted line for CI $(0, 1.19)$ around $y=0.35$, dashed line for CI $(-2.1, 0]$ around $y=-1.25$, and solid line for CI $(-1.19, 0.49)$ around $y=-0.35$. 
 }\label{fig:explained_pp}
\end{figure}

\rev{Our positive-preferring CI has improved sign determination for $q_{1-\alpha+e}\le Y<q_{1-\alpha/2}$ and a more informative lower bound for $0\le Y<q_{1-\alpha+e}$. The price is arguably minimal for applications that care most about interval estimation in the positive direction: $\mathcal I_{\DPp,r}(Y;\alpha)$ is non-positive (rather than entirely negative) whenever $\mathcal I_{\ns}(Y)$ is strictly negative for $-q_{1-e}\le Y<-q_{1-\alpha/2}$, and it is longer than $\mathcal I_{\ns}(Y)$ for $-q_{1-\alpha/2}-q_{1-e}\le Y<-q_{1-\alpha/2}$.}
An example of the positive preferring acceptance regions and the corresponding CIs is shown in Figure \ref{fig:explained_pp}. Another illustration is provided in Figure \ref{fig:all_ci_ar}, which  compares these CIs with those described in  \S~\ref{subsec-review-marginal};  see \S~\ref{subsec-single-comparison} for a detailed discussion.  

 For any true parameter value $\theta>0$, the
exact confidence set obtained by inverting the direction-preferring
acceptance regions,
\[
\mathcal A^{-1}_{\DPp}(Y;r,\alpha)
:=
\left\{
\tau\in\mathbb R:
Y\in\mathcal A_{\DPp}(\tau;r,\alpha)
\right\},
\]
has no greater expected total length than the standard CI, as formalized
in the following proposition.

\begin{proposition}\label{prop:DPvsStandardLength}
Suppose that $\{F_\theta:\theta\in\mathbb R\}$ is a symmetric unimodal
location family with densities satisfying the monotone likelihood ratio
property. Then, for every $\theta>0$,
\[
\mE_{\theta}
\left[
\lambda\left\{
\mathcal A^{-1}_{\DPp}(Y;r,\alpha)
\right\}
\right]
\leq
\mE_{\theta}
\left[
\lambda\left\{
\mathcal I_{\mathrm{std}}(Y;\alpha)
\right\}
\right],
\]
where $\lambda(B)$ denotes the Lebesgue measure of a set
$B\subseteq\mathbb R$.
\end{proposition}
See Appendix \ref{app-proof-prop:DPvsStandardLength} for the proof.



\rev{When applying the direction-preferring CIs, the analyst sets the value of $r$. The derived quantity $\epsilon_{marg}(r,\alpha)$ governs the trade-off between the power to determine the sign and the resulting length of the CI. Specifically, the minimal value of $Y$ for which the sign is declared positive is $q_{1-\alpha+\epsilon_{marg}(r,\alpha)}$. As $r$ increases, $\epsilon_{marg}(r,\alpha)$ decreases, sign determination becomes possible at lower values of $Y$, but at the cost of greater acceptance-region inflation and worse sign determination when $Y<0$. In \S~\ref{sect:choose_r}, we suggest the default $r=1.30$. In all our numerical examples with $\alpha=0.05$, this gives $\epsilon_{marg}(1.30,0.05)=2.82\times10^{-4}$.}



\subsubsection{Direction-preferring Versus Asymmetric Equivariant CIs} \label{sect:comp_equivariant}

When interest lies mainly in interval estimation in one direction, a simpler approach is to construct CIs where the tail error probability is allocated unequally  between the lower and upper bounds. Although the positive sign determination may be the same using both approaches, the simpler approach tends to produce much wider CIs, and does not coincide with the standard CIs for any realized test statistic. In this subsection, we  demonstrate this argument.

The acceptance region using this simpler approach, for positive preference, is $$\mathcal{A}_{eqv}(\theta; \alpha, \epsilon) = \left( \theta - q_{1-\epsilon},\; \theta + q_{1-\alpha+\epsilon} \right),$$
where $\epsilon \in (0, \alpha / 2]$ controls the preference to the positive direction over the negative. 
The resulting CI is \rev{$$ \mathcal{I}_{eqv}(Y;\alpha, \epsilon) = (Y - q_{1-\alpha+\epsilon}, Y + q_{1- \epsilon}).$$}

\begin{figure}
 \centering
\includegraphics[width=0.6\paperwidth, height=9cm]{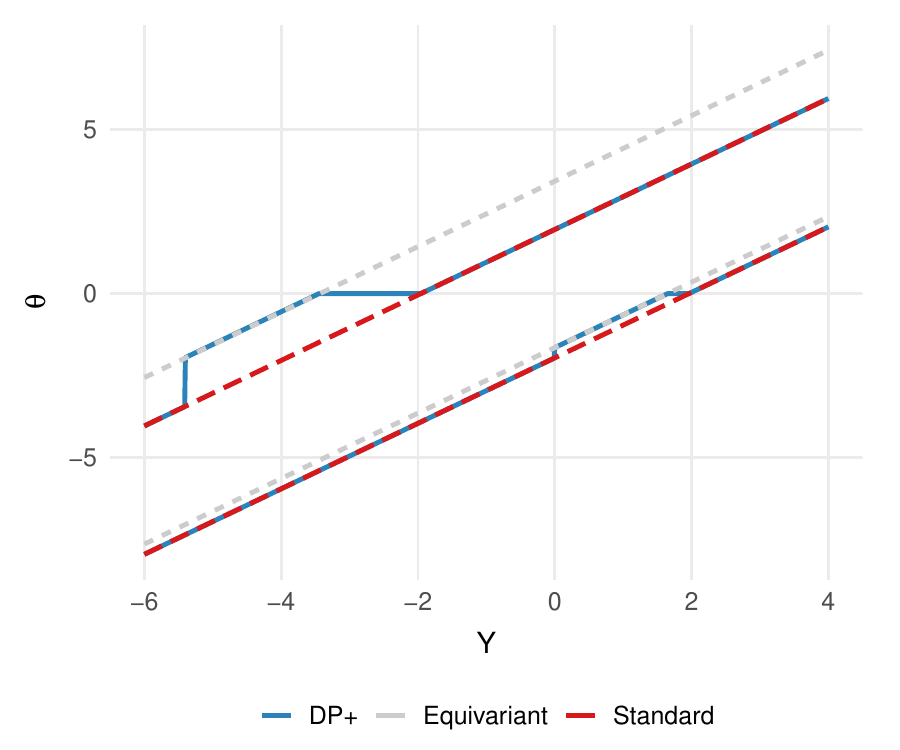}
 \caption{For $Y\sim N(\theta,1)$, comparison of the direction-preferring (positive direction) and asymmetric equivariant 95\% CIs, with \rev{$r=1.30$ and $\epsilon_{marg}(r,0.05)=2.82\times10^{-4}$} for both methods. The upper bound of the 95\% CI is non-positive for $Y < -1.96$ under the direction-preferring method, and for $Y<-3.45$ under the asymmetric equivariant method. The length of the asymmetric equivariant CI is \rev{1.30} times that of the standard CI ($5.08$ vs $3.91$).
 }
 \label{fig:comparing_equi}
\end{figure}


Figure \ref{fig:comparing_equi} provides the three CIs considered $\mathcal I_{std}(Y;\alpha)$, \rev{$\mathcal I_{\DPp,r}(Y;\alpha)$}, and $\mathcal I_{eqv}(Y;\alpha,\epsilon)$. \rev{The equivariant method uses $\epsilon=\epsilon_{marg}(r,\alpha)$}, so the two direction-preferring methods determine the positive sign for the same realized value $Y$, and their acceptance region is the same at the boundary null value $\theta=0$. The direction-preferring CIs have the following advantages over the asymmetric equivariant ones: they tend to be much shorter;
their negative sign determination is considerably better. The only advantage of the asymmetric equivariant CI is a higher lower bound for $ Y > 0$, but this comes at a cost of a much longer CI.


\subsection{The Analysis of Secondary Outcomes in \cite{Manson2019MarineN3VITAL}}\label{subsec-Manson-marginal}

 Direction-preferring CIs are particularly suitable for secondary outcomes in clinical trials. In the study of \cite{Manson2019MarineN3VITAL} introduced in \S~\ref{subsec:intro-motivation},  when fish oil consumption is indeed beneficial, it is important to quantify the magnitude of the reduction in risk. If it is not beneficial, however, the precise quantification is less important.

 For the 22 secondary endpoints in \cite{Manson2019MarineN3VITAL}, five standard $95\%$ CIs exclude the no-effect value, corresponding to two-sided $p$-values at most 0.05.  The test statistic is the estimated log hazard ratio comparing fish-oil supplementation with placebo,  so the no-effect value is zero  (on the original scale of the hazard ratio, the no effect value is one), and the preferred direction is negative (or below one on the original scale). Thus this example is analyzed assuming the direction-preferring CIs are in the negative direction.

\begin{figure}
 \centering
 \centering
 \includegraphics[page=1,width=1\textwidth,height=9cm]{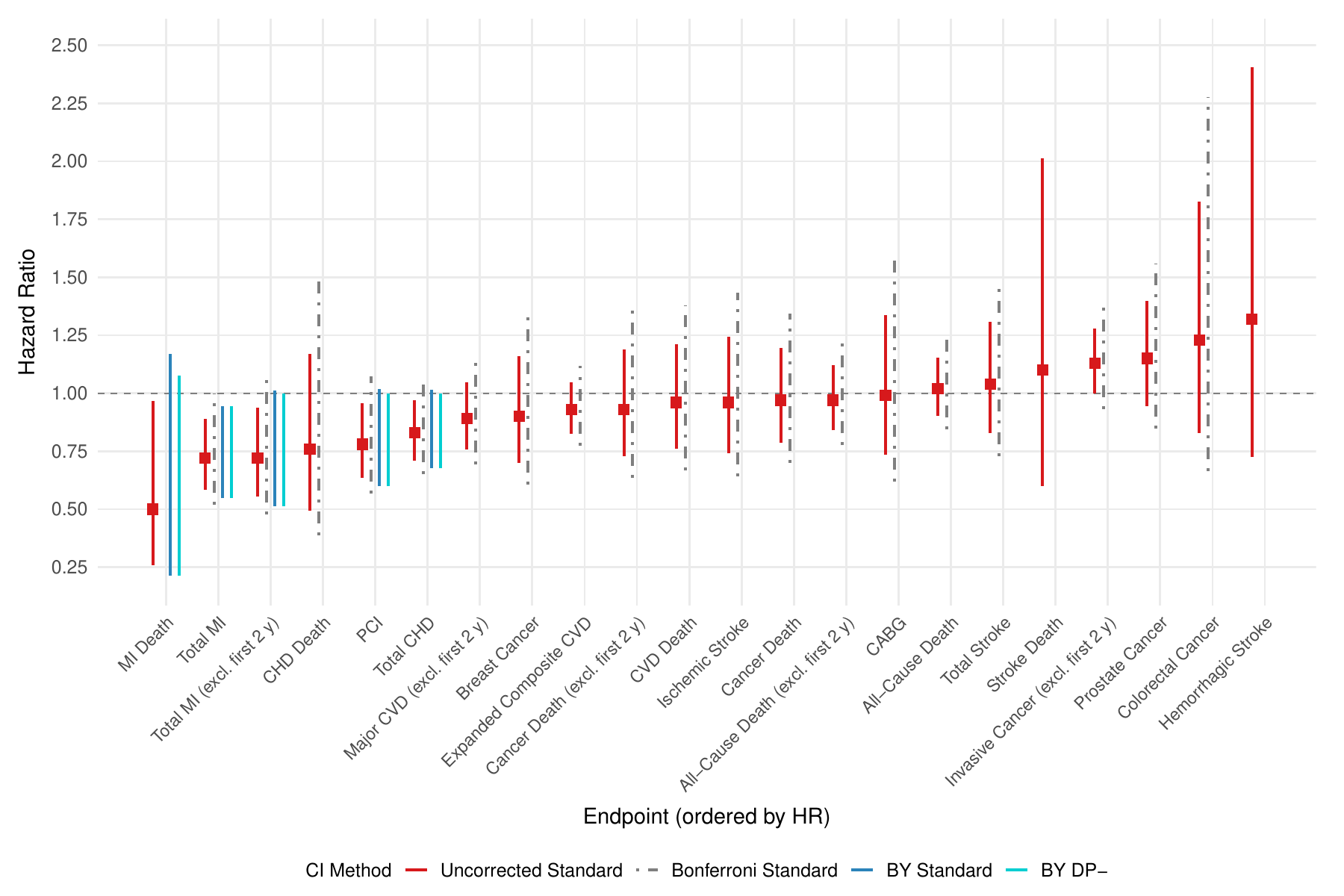}
 \caption{For the 22 secondary endpoints in \cite{Manson2019MarineN3VITAL} and $\alpha=0.05$: the point estimate (red square) along with the standard $1-\alpha$ level CIs (red lines); the Bonferroni-adjusted standard $1-\alpha/22$ level CIs (gray dash dotted lines); the BY-adjusted standard $1-5\alpha/22$ level CIs on the five selected outcomes (solid blue); the BY-adjusted direction-preferring $1-5\alpha/22$ level CIs on the five selected outcomes (solid cyan)\rev{.}} \label{fig:manson}
\end{figure}

 Figure \ref{fig:manson} shows CIs corrected for selection of these five outcomes to guarantee FCR control using the BY procedure \citep{benjamini2005false}, along with the uncorrected standard method (which has no FCR guarantee) and the simultaneous Bonferroni correction method (which may be overly conservative, as it guards against the event that at least one of the 22 constructed CIs fails to cover its parameter value).

 Applying the BY procedure of \cite{benjamini2005false} for the five selected outcomes of interest using standard CIs, only one outcome (total myocardial infarction, CI on the log scale [-0.601, -0.056], on the original scale [0.55, 0.95] ) is entirely \rev{below} the no-effect value. However, using the direction-preferring CIs, four outcomes are entirely \rev{below} the no-effect value: total myocardial infarction (CI on the log scale [-0.601, -0.056], on the original scale [0.55, 0.95]); total myocardial infarction excluding the first 2 years of follow-up (CI on the log scale [-0.668, 0), on the original scale [0.51, 1));
 total coronary heart disease (CI on the log scale [-0.601, 0), on the original scale [0.55, 1)); and PCI (CI on the log scale [-0.388, 0), on the original scale [0.68, 1)).

 Applying the Bonferroni correction, all resulting \rev{CIs} become much wider and cover the null value in Figure \ref{fig:manson}. The loss of sign determination with the simultaneous correction for multiplicity is one of the reasons cited by \cite{harrington2019new} against correcting for multiplicity. We argue that for this application, it is enough to control the FCR at level 0.05 on the selected outcomes of interest, where the selected are the outcomes with nominal 95\% CIs that do not cover the null value (i.e., those with two-sided $p$-values at most 0.05). With FCR rather than simultaneous control, some outcomes can still be identified as beneficial. The direction-preferring intervals strengthen this: the standard two-sided intervals indicate benefit for one outcome, while the direction-preferring intervals indicate benefit for four of the five outcomes.

\section{Generalized CIs Conditional on selection}\label{sec-conditional}
We consider the setting in \cite{weinstein2013selection}, where constructing a CI for $\theta$ is of \rev{interest} only if $|Y|$ exceeds a threshold $c$. Given the selection event \rev{$|Y|>c$}, a {$1-\alpha$ \it conditional CI} satisfies
\begin{equation} \label{eq:cond_ci_general}
    \mathbb{P} ( \theta \in \Ic(Y) \ | \ |Y| > c ) \geq 1 - \alpha.
\end{equation}
If a CI is reported only when the corresponding two-sided test  is rejected  at significance level of $0.05$, the CI construction follows only if \rev{$|Y|>q_{0.975}$} for a normal test statistic $Y\sim N(\theta,1)$. This is the case in the secondary outcomes example in \S~\ref{subsec-Manson-marginal} which we revisit in \S~\ref{subsec-Manson-conditional}.

Since the selection event $\{y: |y|>c\}$ restricts the possible values of $Y$, the distribution is truncated to the region \rev{$\mathcal S_c=(-\infty,-c)\cup(c,\infty)$}. As a result, the usual unconditional coverage of the marginal CI no longer applies. Instead, we construct CIs using the conditional distribution of $Y$ given that $|Y|>c$. Let $F^c_\theta$, $f^c_\theta$, and $q^c_\theta$ be the cumulative distribution, density, and quantile function of this conditional distribution. To ease notation, let $q^c_p=q^c_0(p)$. 

\subsection{Review of conditional CIs}\label{subsec-review-conditional}
The standard and MP CI discussed in the marginal setting, each has a corresponding  conditional CI which is constructed by replacing $f_\theta(y)$ with $f^{c}_\theta$ when obtaining the acceptance region.

The conditional standard CI is obtained by inverting the standard acceptance region   $\Ac_{\ns}(\theta) = \{y: f_\theta^c(y) \geq \xi\}$, where $\xi$ is such that $\mathbb{P} \left( Y \in \Ac_{\ns}(\theta) \ \mid \ |Y|>c\right) = 1 - \alpha$. The only  difference from the marginal construction is that $f_\theta(y)$ is replaced with $f^{c}_\theta(y)$ when defining the acceptance region. Similarly, the MP CI discussed in \S~\ref{subsec-review-marginal} also admits a conditional version detailed in Appendix \ref{sect:description_full}. The resulting conditional CIs are illustrated in Figure \ref{fig:all_ci_ar}, along with the novel conditional direction-preferring CIs we describe in \S~\ref{subsec-conditional-PP}.

\cite{weinstein2013selection} introduced  a third variant, the conditional quasi-conventional CI,  which offers a better balance between early sign determination and length.   Like the conditional CIs discussed above, this variant   treats both directions of inference symmetrically. Since its performance on sign determination and length is  between the above two constructions, we omit it from further comparison for simplicity.

In \S~\ref{sec-single-comparison} we show that the proposed conditional direction-preferring CIs achieve, in the preferred direction for all $\theta>0$, the same sign determination probability as with the conditional MP CI and a shorter average  length than the conditional standard CIs.

\subsection{Direction-preferring conditional CI}\label{subsec-conditional-PP}
\rev{As in the marginal setting in \S~\ref{sect:direction_ci}, we parameterize the conditional direction-preferring CI by the inflation factor $r\geq1$. Let $\tilde{e}=\epsilon_{cond}(r,c,\alpha)\in[0,\alpha/2]$ be the derived conditional tail allocation defined by
\[
q^c_{1-\alpha+\tilde{e}}-q^c_{\tilde{e}}-2c
=r|\Ac_{\ns}(0;\alpha)|.
\]
Thus $r$ is the ratio of the total lengths of the direction-preferring and shortest conditional acceptance regions at $\theta=0$.}

For best positive-sign determination, we  require that when $\theta \leq 0$, the acceptance region has an upper bound no greater than the one at $\theta=0.$ This is satisfied by defining the lower bounds and upper bounds of the acceptance region, $l_1(\theta; r)$ and $ u_1(\theta; r)$, as the solutions to the following two equations:  \begin{enumerate}
 \item \textbf{Length}: \rev{The total length (i.e., the Lebesgue measure) of the acceptance region is
 $\left|\left(l_1(\theta),u_1(\theta)\right)\cap\mathcal S_c\right|=r|\Ac_{\ns}(\theta;\alpha)|$.
 This total length is $u_1(\theta)-l_1(\theta)-2c$ when the acceptance region has two disjoint intervals and $u_1(\theta)-l_1(\theta)$ when it is a single interval.}
 \item \textbf{Coverage probability}: The probability that $Y$ falls within the acceptance region is $1 - \alpha$, i.e.,  $\Fc(l_1(\theta)) + 1 - \Fc(u_1(\theta)) = \alpha$.
\end{enumerate}

We can improve the length in the non-preferred direction by reverting  to the standard acceptance region when $\theta$ is small enough.
The direction-preferring acceptance region (in the positive direction), \rev{$\Ac_{\DPp}(\theta;r,c,\alpha)$}, is therefore as follows:  \begin{equation} \label{eq:spcmp_acceptance}
{\Ac_{\DPp}(\theta;r,c,\alpha)=
\begin{cases}
\begin{aligned}
& \Ac_{\ns}(\theta; \alpha) &\quad \theta > 0 \\
& \left( l_1(\theta; r), u_1(\theta; r) \right) \cap\mathcal S_c & \quad \theta^-_1 < \theta \leq 0 \\
& \Ac_{\ns}(\theta; \alpha) & \quad \theta \leq \theta^-_1 \\
\end{aligned}.
\end{cases}}
\end{equation} \rev{The reversion point $\theta^-_1$ is the negative solution of
\[
F^c_{\theta^-_1}(-c)=1-\alpha/2,
\]
at which the shortest conditional acceptance region is itself entirely negative. Thus reversion to the standard acceptance region is carried out at $\theta\leq \theta^-_1$ because for these values of $\theta$ the acceptance region is   a single negative interval (following a similar logic as in the marginal case).}
\rev{See \S~\ref{sect:choose_r} for details. In all our numerical examples with $\alpha=0.05$ and $c=1.96$, we use $r=1.30$, which gives $\epsilon_{cond}(1.30,1.96,0.05)=1.09\times10^{-3}$.}

By inverting the acceptance region, we obtain the conditional direction-preferring CI; for brevity, its   details are deferred to \cref{sect:pp_ci}. Figure \ref{fig:all_ci_ar} shows the resulting CIs and compares them with those described in \S~\ref{subsec-review-conditional}; see \S~\ref{subsec-single-comparison} for a detailed discussion. %

\subsection{A conditional analysis of secondary outcomes in \cite{Manson2019MarineN3VITAL}}\label{subsec-Manson-conditional}

In \S~\ref{subsec-Manson-marginal} we presented the benefit of using marginal direction-preferring CIs for the 22 secondary outcomes. Since all 22 outcomes were defined in advance, they constitute a family of parameters on which we can apply the BY procedure for FCR control. For comparison, for the five selected outcomes with \rev{marginal} $95\%$ CIs that exclude the null-effect value (i.e., with two-sided $p$-value at most 0.05), we constructed the direction-preferring conditional CIs in order to compare them to the non-conditional ones.

\begin{figure}
 \centering
 \centering
 \includegraphics[page=1,height=7cm]{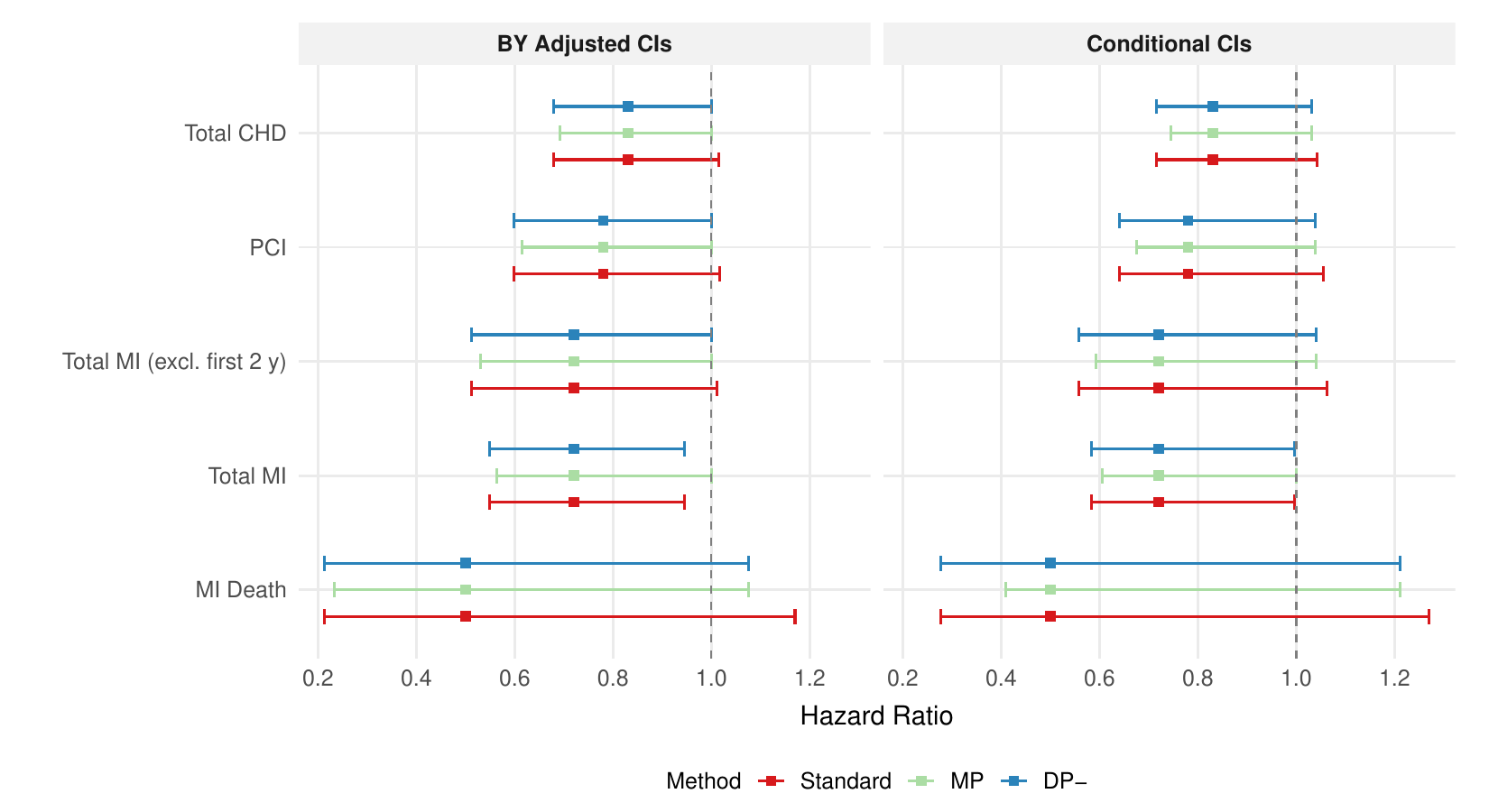}
 \caption{For the five secondary endpoints in \cite{Manson2019MarineN3VITAL} whose 95\% CIs exclude the value of no effect, with $\alpha=0.05$, the BY-adjusted $1-5\alpha/22$ level CIs on the five selected outcomes (left column) and the conditional CIs (right columns) for CI constructed: in the standard 2-sided symmetric fashion (Standard); using modified \rev{Pratt} (MP); using direction-preferring CIs in the negative direction ($\mathrm{DP}^{-}$).}  \label{fig:manson2}
\end{figure}

Figure \ref{fig:manson2} shows the resulting CIs. All of the direction-preferring conditional 95\% CIs include the null-effect value, and  they are considerably longer than the direction-preferring marginal CIs corrected for multiplicity by the BY procedure (which determines a beneficial effect for  four of the five outcomes). We  show in \S~\ref{sect:selected_paramters} that this is indeed expected.

The conditional guarantee is appealing since it is valid regardless of the number of outcomes examined. It also provides a stronger guarantee than FCR control on the family of secondary outcomes, since coordinatewise conditional coverage control implies FCR control \citep{weinstein2013selection}. However, this stronger guarantee comes at the cost of  wider intervals than with the BY procedure. Therefore, we recommend using conditional CIs only when the BY procedure is not appropriate. This is the case, e.g., with outcomes that were not part of the  prespecified family of secondary outcomes. This is also the case when only some of the outcomes are reported in a subsequent publication, as in the case of \cite{Wu2019DietaryFatsCardiometabolic}, which cited as evidence that $n-3$ fatty acid supplementation reduces the risk of  four of the five secondary outcomes whose 95\% CIs exclude the value of no effect, without mentioning the size of the original family of outcomes considered.

\section{Comparison of direction-preferring (marginal and conditional) CIs to standard and MP alternatives
}\label{sec-single-comparison}

The qualitative comparison of the direction-preferring CIs to the standard and MP CIs is similar in the marginal and conditional setting. Therefore, we display it side by side. We compare the shape of the constructed CIs in the $Y$ versus $\theta$ plane in \S~\ref{subsec-single-comparison}, and assess the power to determine the sign and expected length for a range of parameter values in \S~\ref{sect:simulation}.  For brevity, we refer to the positive direction-preferring CI as the $\DPp$ CI.

\subsection{An illustration of the three CI construction methods}\label{subsec-single-comparison}
\begin{figure}
 \centering
 \includegraphics[width=1\textwidth]{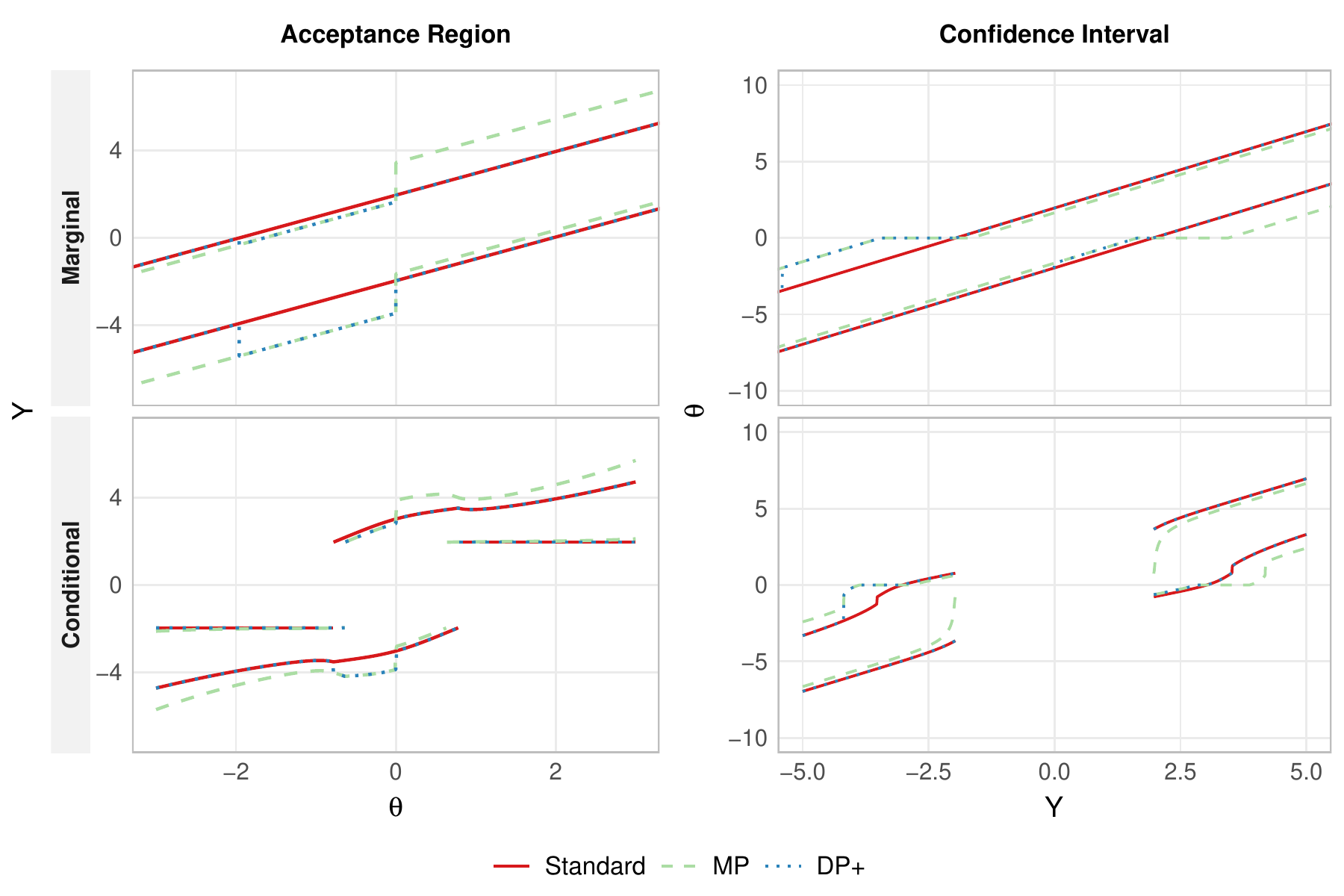}
 \caption{For $Y\sim N(\theta, 1)$, comparison of the acceptance regions and CIs of the \DPp\ with the MP and standard methods at $\alpha =0.05$ and \rev{$r=1.30$}. For the marginal setting (top row), \rev{$\epsilon_{marg}(r,\alpha)=2.82\times10^{-4}$}. For the conditional setting (bottom row), the truncation value is $c=1.96$, and \rev{$\epsilon_{cond}(r,c,\alpha)=1.09\times10^{-3}$}.
 }
 \label{fig:all_ci_ar}
\end{figure}

A comparison of the CIs and acceptance regions in the marginal and conditional settings is shown in Figure \ref{fig:all_ci_ar}, where $Y \sim N(\theta, 1)$.  The \DPp CI can determine that $\theta > 0$ for much lower realizations of the test statistic than the other two methods. Specifically, the  CI is entirely above zero when $Y$ is greater than:  1.65 (marginal setting) and 2.81 (conditional setting) for $\DPp$ CI; 1.96 (marginal setting) and 3.02 (conditional setting) for the standard  and MP CIs. Moreover,  the \DPp CI determines that $\theta \leq 0$ for the same observed values as the standard CI (although the standard CI can assert that $\theta < 0$), i.e., when  $Y$ is smaller than:  -1.96 (marginal setting) and -3.02 (conditional setting). The smallest Y above which the CI lower bound is $>0$ is also the same for the \DPp and standard CIs, and it is much higher (i.e., less informative) for the MP CI. Specifically, the lower bound is $>0$ when $Y$ is greater than: 1.96 (marginal setting) and 3.02 (conditional setting) for  \DPp and standard CIs; 3.45 (marginal setting) and 3.87 (conditional setting) for the MP CI.
Finally, the largest $Y$ below which the CI upper bound is $<0$ is the same for the \DPp and MP CIs, and it is much higher (i.e., more informative) for the standard CI. Specifically, the upper bound is $<0$ when $Y$ is less than: -3.45 (marginal setting) and \rev{-3.87} (conditional setting) for \DPp and MP CIs; -1.96 (marginal setting) and -3.02 (conditional setting) for the \rev{standard} CI.

\begin{remark}\label{rem-lengthMP}
The MP CIs can have a shorter length compared to the standard CI when $Y$ is close to $0$.
For small $|Y|$, the CI length is determined by the upper bound of an acceptance region obtained for $\theta < 0$, and by the lower bound of an acceptance region for $\theta > 0$ (see Figure \ref{fig:all_ci_ar}).
As both acceptance region bounds are closer to $0$ compared to those of the shortest (standard) acceptance region, after the inversion, the resulting CI is shorter than the standard CI.
This happens only when zero is included in the CI, so this does not aid in the determination of the direction of the effect. As $|Y|$ grows larger, and the CI results in the inversion of acceptance regions from only one side of zero, the MP CI becomes longer than the standard CI.
\end{remark}

\subsection{Simulations comparing \DPp with MP and standard CIs} \label{sect:simulation}

By construction, the direction-preferring CIs will have better sign determination for the preferred sign compared to the standard and MP CIs. The goal of the simulation is to quantify the increase in sign determination, as well as the length and the minimal effect estimate.
We adopt Tukey's viewpoint that, in practice $\theta \neq 0$ \citep{tukey1991philosophy}. Consequently, our focus is on \rev{sign determination}, with or without the exclusion of 0.
The test statistic $Y$ is sampled from $N(\theta_i, 1)$ where $\theta_i \in \{-4, -3.95, \ldots, 4\}$. In the conditional setting samples are drawn from the truncated normal distribution $[Y| |Y| > 1.96]$.

Three metrics are compared: the average length of the CI; the average minimal effect estimate (for $y > 0$ it is the average of the CI lower bound, and for $y < 0$ it is the average of the CI upper bound); and the probability of sign determination from the CIs (i.e., the probability that the interval includes only one sign).

\begin{figure}
 \centering
  \includegraphics[width=\textwidth, height=0.5\paperheight]{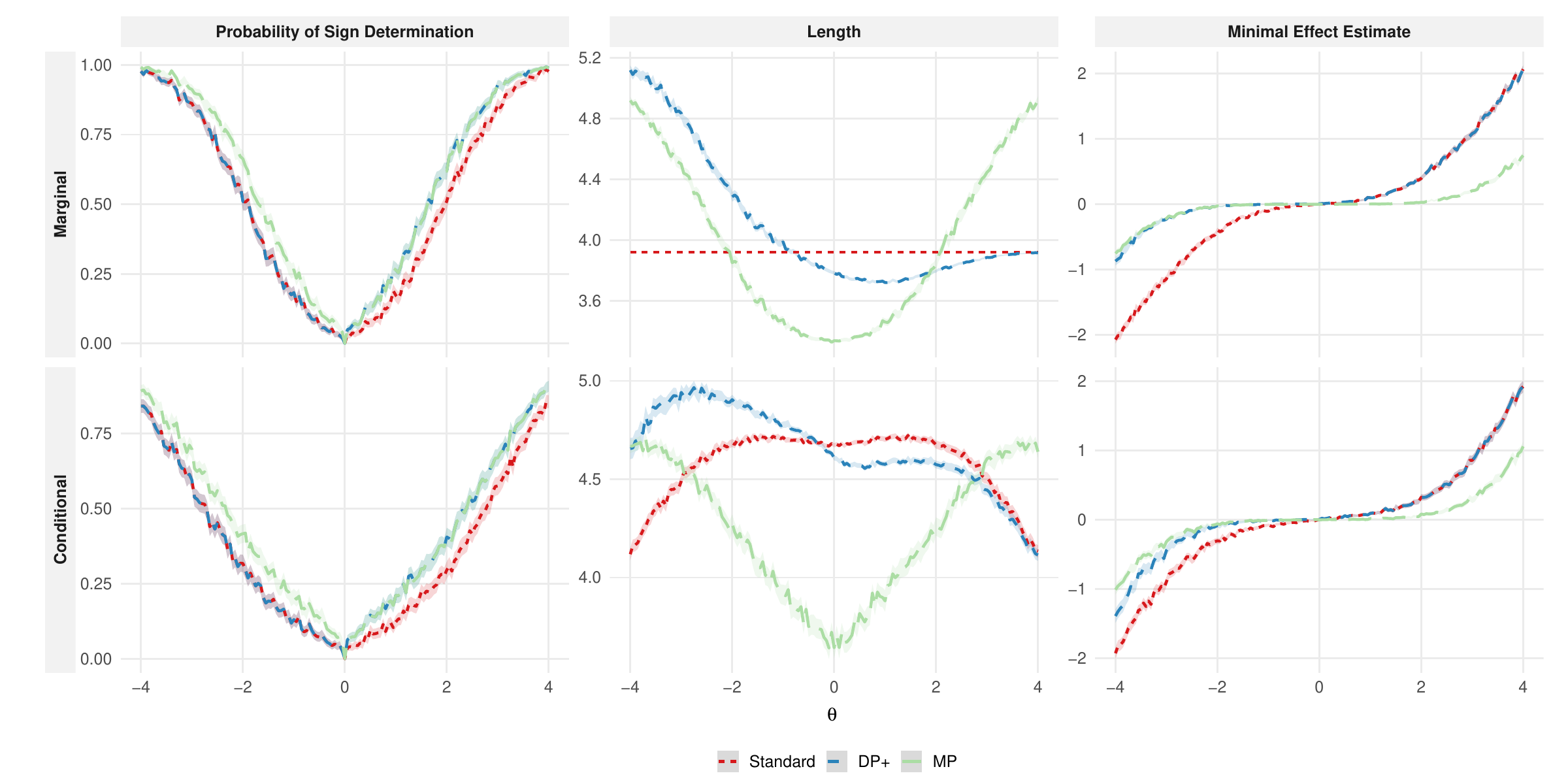}
 \caption{Comparing the methods for constructing CIs with $\alpha=0.05$ as a function of $\theta$ (see \S~\ref{sect:simulation} for data generation process). The x-axis is various values of $\theta$. \rev{For both the marginal and conditional methods, $r=1.30$; the derived allocations are $\epsilon_{marg}(r,\alpha)=2.82\times10^{-4}$ and $\epsilon_{cond}(r,c,\alpha)=1.09\times10^{-3}$, respectively.} To make the comparison fair the MP CI (marginal and conditional) is also set at \rev{$r=1.30$}. Results are based on \rev{1000} data generations for every $\theta$. }
 \label{fig:sim_summary}
\end{figure}

Figure \ref{fig:sim_summary} shows the results. The qualitative conclusions for both conditional and marginal settings are similar. As expected, the \DPp CIs have for $\theta >0$ : a better  sign determination probability compared to the standard CI;
the same sign determination probability as the MP CIs, but with a much higher (more informative) lower bound (minimal effect estimate) on average. On the other hand, for $\theta < 0$, the \DPp CIs have: a higher (less informative) average minimal effect estimate than the standard CIs, with the same sign determination probability; worse sign determination than the MP CIs, with slightly less informative minimal effect estimation on average.

Since the \DPp acceptance regions revert to the standard possible ones for sufficiently large $\theta$, for large estimator values, the \DPp CIs are identical to the standard CIs. This means that on average, for large values of $|Y|$ the \DPp CIs are shorter than the MP CIs. For smaller values of $|Y|$ the MP CI is shorter than the standard CI (see Remark \ref{rem-lengthMP} for an explanation).

\section{Comparing BY-Adjusted and Conditional CIs} \label{sect:selected_paramters}

Conditional coverage control entails FCR control for independent test statistics \citep{weinstein2013selection}. This invites a comparison with the CIs designed to control the FCR, such as the BY- adjusted CIs. An important difference is that the conditional control applies to each specific parameter individually, whereas the control of  FCR is based on average expectation across all parameters. So the conditional coverage is controlled regardless of the number of parameters $m$, as long as the selection rules is based on individual thresholding, i.e., selection rules such as $|Y_i| > t_i$.

 For selected CIs, we cannot prove FCR control for general dependence across the estimators, but  the conditional CIs control a closely related error, the marginal FCR,  over the selected parameters.
The result is formalized in the following proposition.

\begin{proposition} \label{prop:mfcr}
    Assume that \rev{for every $i$ with $\mathbb{P}(i\in S(\boldsymbol{Y}))>0$}, the conditional CI can be constructed at level $\alpha$ to satisfy \rev{$\mathbb{P}\{\theta_i\in\mathcal I^c(Y_i)\mid i\in S(\boldsymbol{Y})\}\geq1-\alpha$}. Assume also that the probability of selecting at least one parameter is positive, i.e.,  $\{i: \mathbb{P}(i \in S(\boldsymbol{Y})) > 0\} \neq \emptyset.$ Then
    \begin{equation} \label{eq:mfcr}
\mathrm{mFCR}\rev{:=}  \frac{\mathbb{E}  \left( \sum_{i=1}^m\mI(\theta_i \notin \mathcal{I}^c(Y_i), i \in S(\boldsymbol{Y})) \right) }{{ \mathbb{E} \left(\sum_{i=1}^m\mI(i \in S(\boldsymbol{Y})) \right) } }\leq \alpha.
\end{equation}
\end{proposition}

\begin{proof}
  The denominator of the mFCR is positive, since  $\mathbb{E} \left(\sum_{i=1}^m\mI(i \in S(\boldsymbol{Y})) \right)\geq \mP(i^* \in S(\boldsymbol{Y}))>0$ for $i^*\in \{i: \mathbb{P}(i \in S(\boldsymbol{Y})) > 0\}$. Therefore,
  \begin{eqnarray}
		&& \mathrm{mFCR}  = \frac{\sum_{i=1}^m \mathbb{E}  \left( \mI(\theta_i \notin \mathcal{I}^c(Y_i), i \in S(\boldsymbol{Y})) \right) }{{\sum_{i=1}^m \mathbb{E} \left(\mI(i \in S(\boldsymbol{Y})) \right) } } \nonumber \\ && =
			 \frac{\sum_{i=1}^m \mathbb{P} (i \in S(\boldsymbol{Y})) \mathbb{P} (\theta_i \notin\mathcal{I}^c(Y_i) \mid i \in S(\boldsymbol{Y}))}{\sum_{i=1}^m \mathbb{P} (i \in S(\boldsymbol{Y})) }  \leq
			 \frac{\sum_{i=1}^m \mathbb{P} (i \in S(\boldsymbol{Y})) \alpha }{\sum_{i=1}^m \mathbb{P} (i \in S(\boldsymbol{Y})) }  = \alpha, \nonumber
		\end{eqnarray}
	    where the inequality follows from  the assumption of the proposition. 
\end{proof}



Constructing conditional CIs requires obtaining the conditional distribution, while existing marginal CIs, under certain assumptions, can be modified to control the FCR by adjusting the coverage probability. Once the CIs are obtained, conditional CIs offer greater flexibility than BY-adjusted CIs. Conditional CIs maintain coordinatewise conditional coverage and mFCR control for any further subset of selected parameters chosen without using the observed values of $\boldsymbol{Y}$. If the components of $\boldsymbol{Y}$ are independent, they also control the FCR. In contrast, the FCR guarantee of BY-adjusted CIs applies to the set of selected parameters for which the adjustment was made and would not necessarily be preserved after further selection.


\begin{figure}
 \centering
\includegraphics[width=0.7\paperwidth]{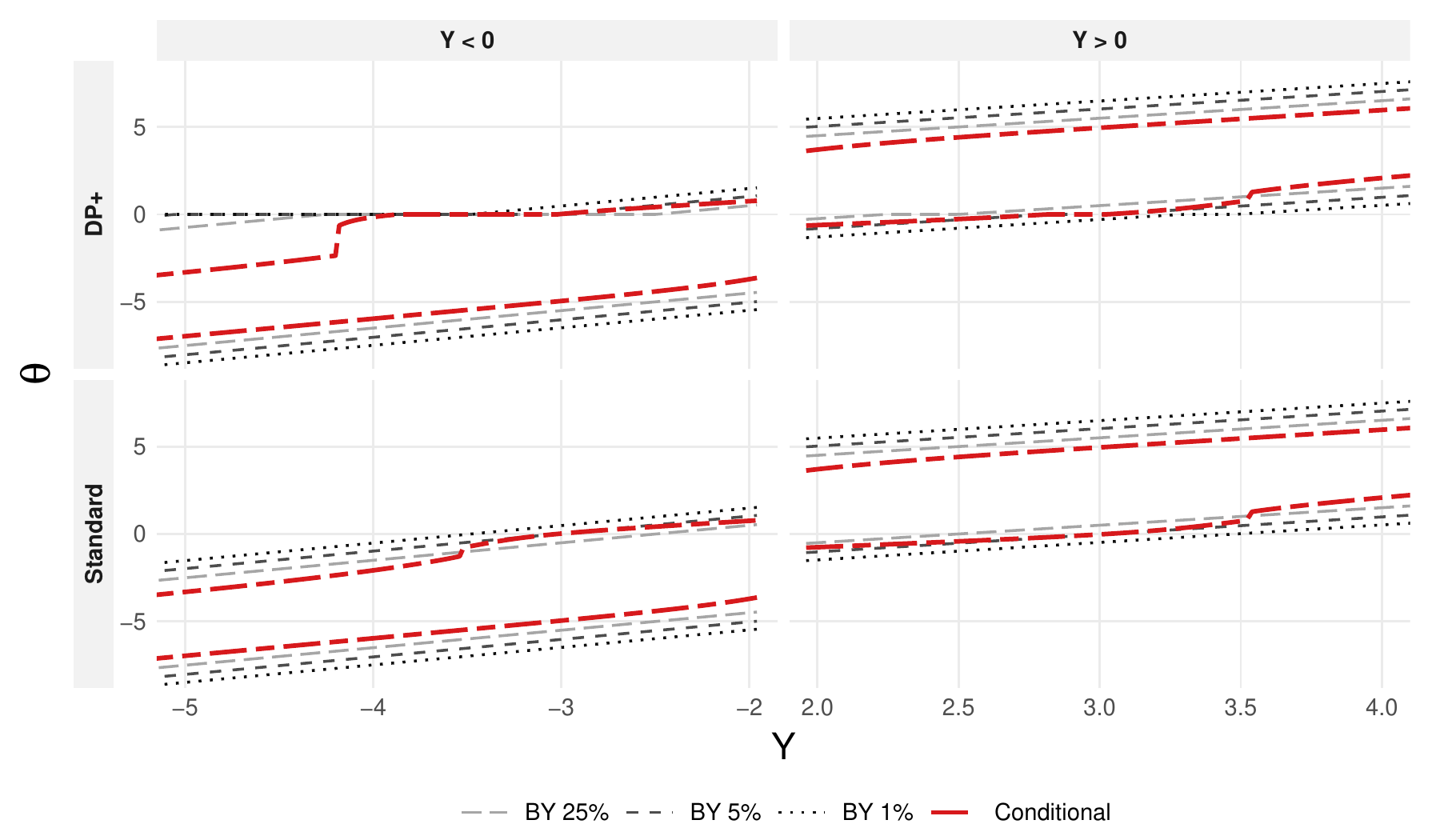}
 \caption{Comparing the BY-adjusted CI when the fraction of selected parameters is $0.25$, $0.05$, or $0.01$ with the conditional \DPp and conditional standard CIs. The test statistics are $Y_i\sim N(\theta,1)$, the truncation value is $c=1.96$, $\alpha=0.05$, and \rev{$r=1.30$; the derived allocations are $\epsilon_{marg}(r,\alpha)=2.82\times10^{-4}$ and $\epsilon_{cond}(r,c,\alpha)=1.09\times10^{-3}$}. }
 \label{fig:comparing_fcr_and_conditional}
\end{figure}

Figure \ref{fig:comparing_fcr_and_conditional} compares the conditional standard and direction-preferring CIs with the BY-adjusted standard CI, $\mathcal{I}_{\ns}(.; \frac{|S(\boldsymbol{Y})|}{m} \alpha )$. This comparison includes BY-adjusted CIs for proportions of selected parameters \rev{$\frac{|S(\boldsymbol{Y})|}{m}\in\{1\%,5\%,25\%\}$} and juxtaposes them with the direction-preferring and the standard conditional CIs, both using a selection threshold of $c=1.96$. The estimator, $Y$, is sampled from $N(\theta, 1)$ and all CIs are constructed at $0.95$ confidence level.

Our analysis reveals that the standard conditional CI performs comparably to the BY-adjusted CIs for sign determination when only 5\% of parameters are selected.

Additionally, we observe that the bounds farthest from zero are consistently closer to zero for conditional CIs compared to BY CIs (an improved upper bound for the BY CIs is suggested in  \citet{tzviel2025benjamini}, see \S~\ref{sect:disc} for a brief discussion). The bound closest to zero is comparable between the standard conditional CI and the BY-adjusted CI with 5\% parameters selected for small $|Y|$. For $Y < 0$, when $Y$ is relatively large, the conditional direction-preferring upper-bound is relatively large. However, as $|Y|$ increase, the conditional CIs converge towards the standard marginal CIs without adjustment. The BY-adjusted CIs are of fixed length as function of $Y$, and increase as $\frac{|S(\boldsymbol{Y})|}{m}$ decreases, so their length coincides with the marginal standard CI only when all parameters are selected.

An interesting question that arises from the preceding analysis is how conditional CIs compare with their  BY-adjusted CI counterparts in terms of sign determination. Any non-equivariant CIs can be adjusted to control the FCR using the selected fraction $\frac{|S(\boldsymbol{Y})|}{m}$ in the confidence level. The following \rev{theorem} shows that, when the selected fraction is at least as large as the probability of selection when the null hypothesis is true, the BY-adjusted CIs have greater power for sign determination than the corresponding conditional CIs.

\begin{theorem}\label{thm:SD-Conditional-BY}
    Let $Y_i, \; i = 1, \ldots, m$ be the test statistics for parameters $\theta_i, \; i = 1, \ldots, m$. \rev{Let $0<\alpha<1/2$.} Assume
 \begin{enumerate}
   \item Selection is based on a fixed threshold: $S(\boldsymbol{Y}) =
   \{i: |Y_i| > c\}$\rev{, where $c>0$}.
   \item $Y_i$ comes from a \rev{continuous, symmetric} unimodal location family, i.e., $Y_i\sim F_{\theta_i}$, with distribution function
   \[
   F_{\theta_i}(y)=F(y-\theta_i),
   \]
   where $F$ is the baseline unimodal distribution centered at 0\rev{, with a positive density $f$ that is nonincreasing on $[0,\infty)$}.
   \item \rev{The marginal direction-preferring CI family is nested in its error level.}
 \end{enumerate}
Let \rev{$\mathcal I_{\DPp,r}(Y_i;\alpha)$ and $\mathcal I^c_{\DPp,r}(Y_i;\alpha)$} denote the marginal and conditional CIs, respectively, at level $\alpha$, where $r\in[1,\infty)$ is the inflation ratio between the acceptance region length of the (marginal or conditional) direction-preferring CIs and standard CIs \rev{at $\theta=0$}. Recall that $r$ is a tuning parameter set by the analyst, and $r=1$ corresponds to the standard construction of marginal and conditional CIs.
Let $t_0=\mP_{\theta_i=0}(|Y_i|>c)=2(1-F(c))$ be the probability of selecting a test statistic when the parameter value is zero, and \rev{let $\widehat{\pi}_m=|S(\boldsymbol{Y})|/m$}. Then if the fraction selected is at least $t_0$, i.e., \rev{$\widehat{\pi}_m\geq t_0$}, we have better sign determination in the preferred direction with the BY-adjusted CIs, i.e., for $Y_i>0$: if \rev{$0\notin\mathcal I^c_{\DPp,r}(Y_i;\alpha)$, then $0\notin\mathcal I_{\DPp,r}(Y_i;\alpha\widehat{\pi}_m)$}.
\end{theorem}
See Appendix \ref{app-proof-thm:SD-Conditional-BY} for the proof, and Remark \ref{rem-2sided-SD-advantage-BY} for a similar advantage of BY-adjusted CIs over conditional CIs for standard (symmetric) CIs.

\rev{The sufficient condition holds for every realization with at least one selection whenever $t_0\leq1/m$. For the asymptotic probability comparison, it is enough that the selected fraction converge in probability to a limit no smaller than $t_0$, provided that the marginal sign-determination threshold is continuous in the error level. This includes the boundary case in which the selected fraction converges to $t_0$, even though its probability of exceeding $t_0$ need not converge to one. We thus have the following corollary; see Appendix \ref{app-proof-cor:SD-Conditional-BY} for the proof.}

\begin{corollary}\label{cor:SD-Conditional-BY}
    {Under the assumptions of Theorem \ref{thm:SD-Conditional-BY}, define
    \[
    \widehat{\pi}_m=\frac{|S(\boldsymbol{Y})|}{m},
    \qquad
    \begin{aligned}
    A_m&=\{0\notin\mathcal I^c_{\DPp,r}(Y_i;\alpha),\ Y_i>0\},\\
    B_m&=\{0\notin\mathcal I_{\DPp,r}(Y_i;\alpha\widehat{\pi}_m),\ Y_i>0\}.
    \end{aligned}
    \]
    Suppose $\widehat{\pi}_m\xrightarrow{p}\pi_S$ for a constant $\pi_S\geq t_0$, where $\pi_S$ denotes the limiting overall selected fraction, and that the upper endpoint at $\theta=0$ of the marginal direction-preferring acceptance region is continuous in its error level at $\alpha\pi_S$. Then
    \[
    \liminf_{m\to\infty}\mP(A_m)\leq\liminf_{m\to\infty}\mP(B_m).
    \]}
\end{corollary}

\begin{remark}\label{rem-2sided-SD-advantage-BY}
    Consider the special case of standard two-sided CIs, obtained by setting \rev{$r=1$} in equation \eqref{eq:pp-marginal} or \eqref{eq:spcmp_acceptance}. In this case, an acceptance-region comparison similar to that used in the proof of Theorem \ref{thm:SD-Conditional-BY} implies an analogous statement for sign determination (via exclusion of $0$):
    \begin{itemize}
        \item If the fraction selected is at least \rev{$t_0$}, i.e., $|S(\boldsymbol{Y})|/m\geq t_0$, then $0\notin \mathcal I^c_{std}(Y_i;\alpha)$ implies $0\notin \mathcal I_{std}(Y_i;\alpha |S(\boldsymbol{Y})|/m)$.
        \item If \rev{$|S(\boldsymbol{Y})|/m\xrightarrow{p}\pi_S\geq t_0$, where $\pi_S$ is the limiting overall selected fraction, and the marginal sign-determination threshold is continuous at error level $\alpha\pi_S$}, the analogous asymptotic probability comparison follows.
    \end{itemize}
\end{remark}




\subsection{Simulation study}

We compare four properties of CIs constructed after selection:
(i) interval length,
(ii) the marginal false coverage rate (mFCR, defined in \cref{eq:mfcr}),
(iii) the minimal effect size, defined as the smallest value in the CI consistent with the estimated sign,
and (iv) the proportion of parameters for which the CI determines a sign (i.e., it is a subset of \rev{$[0,\infty)$ or $(-\infty,0]$}).
We compare three CI procedures—MP, \DPp, and standard—each constructed either conditionally on selection
or using marginal BY adjustment.
The BY-adjusted CIs use confidence level
$1 - \frac{|S(\boldsymbol{Y})|}{m}\alpha$.

To study these properties, we simulate data $Y_i \sim N(\theta_i, 1)$ for $i = 1,\ldots,p$, with $p=20$.
The proportion of non-zero parameters $\theta_i$ takes values $0.05$ and $0.5$,
representing sparse and dense regimes, respectively. Additional settings  with larger $p$ and more proportions are deferred to \Cref{sect:additional_fcr_sims}.
All non-zero $\theta_i$ share a common effect size
\rev{$\theta\in\{-3.00,-1.50,-0.50,0.00,0.50,1.50,3.00\}$. Each configuration uses 1000 simulation replicates.}
Dependence is introduced using an AR(1) correlation structure with
$\Sigma_{ij} = 0.7^{|i-j|}$.
Selection is performed using the rule $|Y_i| > 1.96$,
corresponding to a nominal two-sided 5\% threshold.

\begin{figure}
 \centering

 \par\medskip\includegraphics[width=0.7\paperwidth]{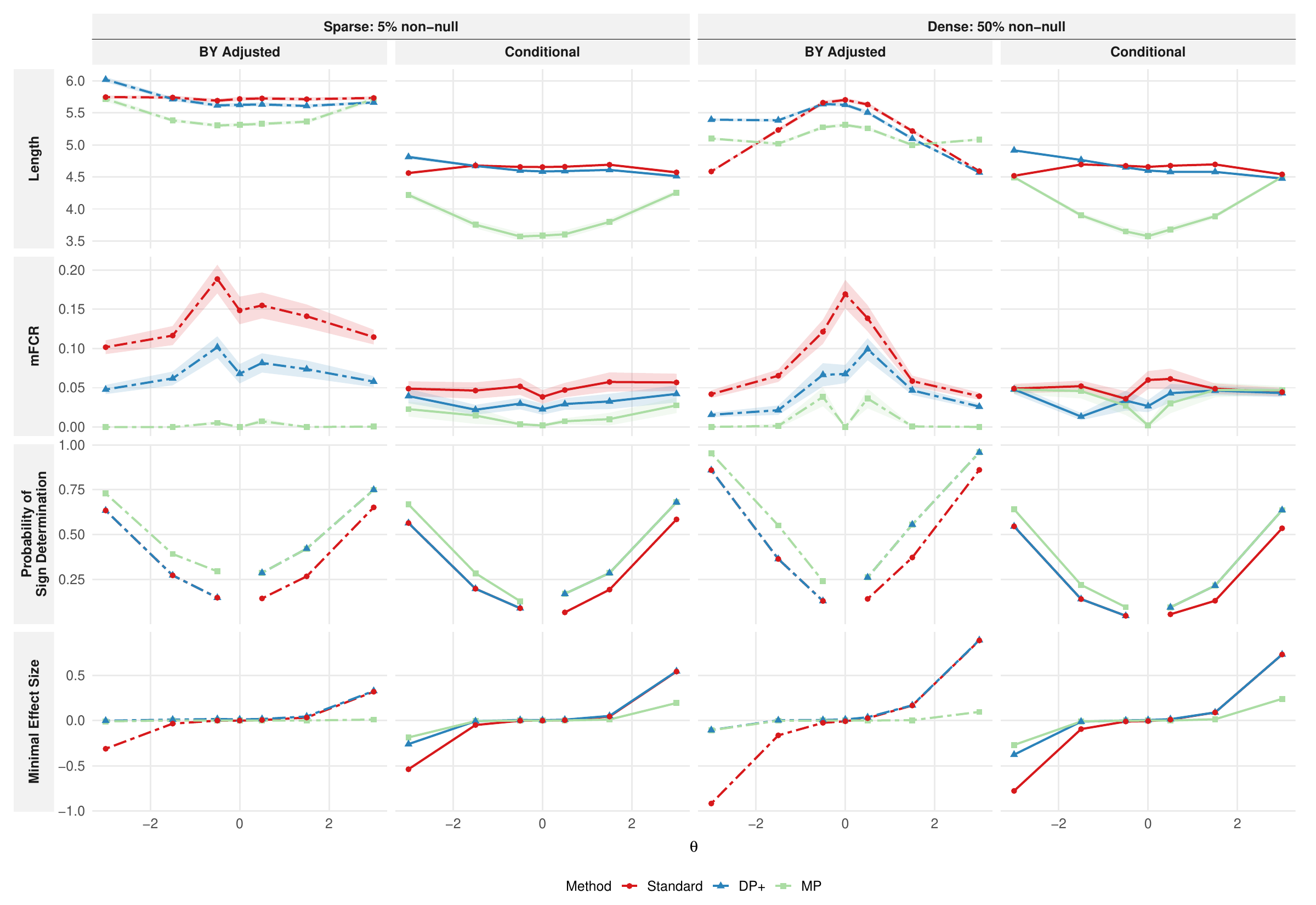}
 \caption{Comparison of conditional and BY-adjusted CIs at nominal 95\% coverage.
Data are generated as $Y_i \sim N(\theta_i,1)$ with AR(1) dependence
$Cor(Y_i,Y_j)=0.7^{|i-j|}$.
Sparse and dense settings are in the two left and two right columns, respectively. The x-axis shows the common non-zero effect size $\theta$.
}
 \label{fig:comparing_fcr_and_conditional_simulation}
\end{figure}

\Cref{fig:comparing_fcr_and_conditional_simulation} summarizes the results.
First, across almost all configurations, the conditional CIs are shorter than their BY-adjusted counterparts.
This is due to the fact that, when the observed estimate is sufficiently large, the conditional CI converges to the corresponding marginal CI,
whereas the BY-adjusted CI remains inflated through its dependence on the proportion of selected parameters.
As a result, BY-adjusted CIs may be longer even though they  produce sign determinations more frequently.

Second, conditional CIs control the mFCR across all configurations
(see \cref{prop:mfcr}),
whereas BY-adjusted CIs may substantially exceed the nominal level,
particularly in sparse regimes and for weak signals.
In some settings, the mFCR of the BY procedure reaches values up to three times the nominal level.

While BY-adjusted CIs may fail to control the mFCR,
they do control the FCR in all settings considered, see  \Cref{sect:additional_fcr_sims}.

Finally, the behavior of the minimal effect size depends strongly on the regime.
In the sparse setting, conditional CIs tend to yield larger minimal effect sizes,
reflecting stronger conditional evidence once selection occurs.
In contrast, in denser settings the BY-adjusted CIs become competitive and may yield larger minimal effect sizes,
consistent with the reduced impact of the multiplicity adjustment.

\section{Application to the  SURPASS-CVOT trial }
\label{sect:real_example}
The SURPASS-CVOT trial compared  major cardiovascular outcomes of Tirzepatide and Dulaglutide in participants with Type 2 Diabetes \citep{Nicholls2025Cardiovascular}. It is an active-comparator-controlled, double-blind, trial for noninferiority with 13,299 patients with Type 2 diabetes and established atherosclerotic cardiovascular disease. The primary endpoint is a composite of death from cardiovascular causes, myocardial infarction, or stroke. The secondary endpoints included individual components of the primary composite among other factors. They were analyzed either by \rev{Cox} proportional hazard models or by differences of proportions.
A graphical testing approach was prespecified to evaluate sequentially the primary endpoint noninferiority, and then its superiority and the secondary endpoints' superiority while preserving a level  5\% FWER.

\begin{figure}
 \centering

 \par\medskip\includegraphics[width=0.7\paperwidth]{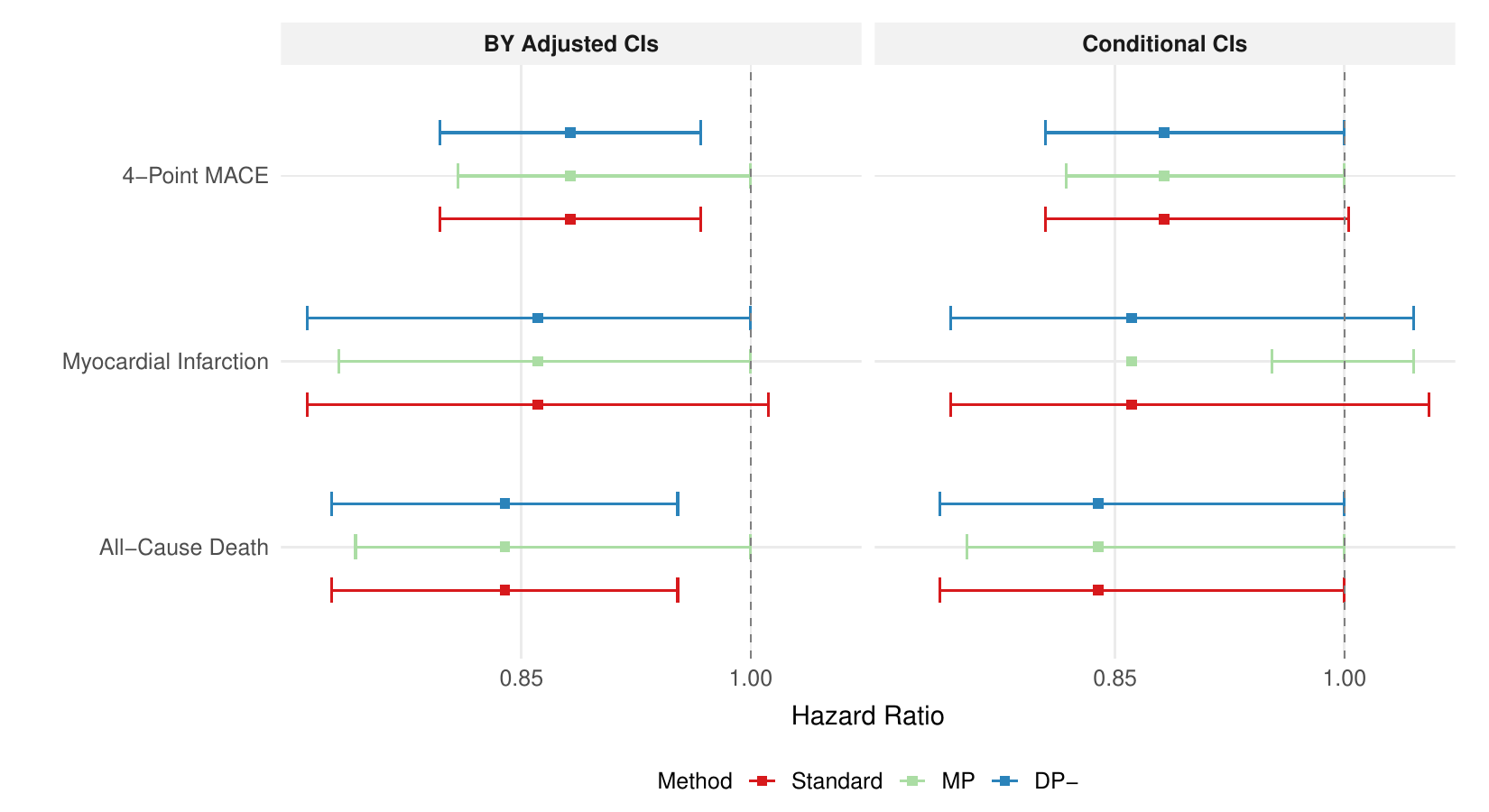}
 \caption{\rev{Eight of the 12 secondary endpoints in \cite{Nicholls2025Cardiovascular} have standard 95\% CIs that do not cover the no-effect value}, and we display here the three that have hazard ratios as endpoints. The BY-adjusted $1-8\alpha/12$ level CIs on the three selected outcomes (left column) and the conditional CIs (right columns) are constructed in the standard two-sided symmetric fashion (Standard), using modified \rev{Pratt} (MP), or using direction-preferring CIs in the negative direction ($\mathrm{DP}^{-}$). Note that the myocardial infarction endpoint has a test-statistic value of approximately $-1.964$, very close to the selection threshold $-1.96$; the conditional MP CI therefore excludes the observed value.}
 \label{fig:surpass_example}
\end{figure}

Tirzepatide was found noninferior to dulaglutide (hazard ratio, 0.92; 95.3\% CI, 0.83 to 1.01; $P=0.003$ for noninferiority) but did not meet the criterion for superiority ($P=0.09$). Hence all results for the secondary endpoints were reported using point estimates and 95\% CIs. The trial authors explicitly noted that CIs for secondary endpoints were not adjusted for multiplicity and should be used for exploratory purposes rather than definitive hypothesis testing. It is, however, hard to avoid noticing that some of them do not cover the no-effect value. Specifically, three hazard ratios out of the six secondary outcomes that are hazard ratios have 95\% CIs that do not cover the no-effect value. Moreover, five of the six mean differences have 95\% CIs that do not cover the no-effect value. Figure \ref{fig:surpass_example} shows the CIs for the three secondary outcomes that are hazard ratios, out of the eight selected among the 12 secondary outcomes. By construction, the upper bound of the direction-preferring CIs in the negative direction is at most as large as the upper bounds from the other two methods. \rev{The advantage over the standard method is particularly apparent for myocardial infarction: the BY-adjusted standard CI includes one, 
 and the $\mathrm{DP}^{-}$ CI excludes one. 
 } The remaining five endpoints had very strong signal; see Appendix \ref{app-sect:real_example}.

\section{Discussion} \label{sect:disc}



CIs are commonly used with hypothesis testing. Unlike CIs, hypothesis testing can indicate the direction or sign of a parameter but fails to convey the effect size and associated uncertainty in parameter estimation. This problem is exacerbated by 'selective inference,' which occurs when only certain parameters are selected for highlighting among the reported. Considering the importance of adjusting CIs for selection, the question arises: which selection adjustment approach should be employed? These methods can be evaluated based on their error guarantee, flexibility and sign-determination capability. For a prespecified pool of inferences, such as the set of primary endpoints or the set of predetermined secondary endpoints all three approaches offer guarantees: the simultaneous method offers FWER control, the BY intervals offer FCR control, and the conditional intervals offer conditional coverage per each interval as well as mFCR control over the selected. In terms of sign determination the simultaneous multiplicity adjustment has the lowest power, conditional CIs have more power and the most powerful are the BY CIs. However, simultaneous and conditional CIs hold their guarantees even when a subset is further selected. Conditional CIs  can be used even on a set which was not prespecified, such as exploratory outcomes in clinical trials, as long as the selection rule was prespecified.

An important consideration in construction of CIs is the sign-preference. Analysts can construct direction-preferring CIs by allocating the coverage probability $\alpha$ differently between the ends of the CIs. While this approach yields better minimum effect size estimation for parameters of the preferred sign, it can significantly reduce sign-determination power for the non-preferred sign. The non-equivariant direction-preferring CIs suggested in this work strike a better balance between these two needs. For the non-preferred sign,  sign determination is barely affected, while the estimation of the minimal effect size is reduced. However, the preferred sign benefits from higher sign-determination power and improved minimal effect size estimation.

Direction-preferring CIs may be particularly useful in secondary endpoint analysis in clinical trials where the analyst's interest lies in detecting additional beneficial endpoints. They also allow identification of potentially harmful endpoints, although they may provide less precise estimates of the extent of harm. Importantly, in clinical trials, it is common to report only a subset of secondary endpoints in subsequent studies. In such cases, using the conditional CI approach is recommended, as it ensures that statistical guarantees, such as coverage probability or FCR control, remain valid for the selected subset of parameters.

\rev{For the conditional CIs considered here, selection is two-sided; for example, selection may occur when a symmetric 95\% CI excludes the value of no effect.} \cite{tzviel2025benjamini} suggested an improvement for the selected parameters when the selection is based on passing a threshold in one direction, say only endpoints showing benefit to some degree.
Their intervals are similar to those proposed by Benjamini and Yekutieli (2005) on the inward end, and to the
marginal unadjusted CIs on the outward end. The center of the suggested CI is a shrunk estimator of the selected parameter. This simple improvement is uniformly better for the one-directional selection, and they suggest how to apply it for a two-directional selection. They \rev{prove FCR control} for independent estimators and provide simulation evidence for its robustness under dependency.

For conditional CIs, we only considered selection when $|Y|$ is large.
In Appendix \ref{sect:conditional_os} we address the problem of selection of the form $Y>c$, where the resulting conditional CIs may become unattractively wide.
Conditional CIs have been studied for more general selection events \citep{fithian2014optimal}. As long as the conditional distribution of $Y$ is known, the direction-preferring approach can be applied. Extending this approach to other selection events is left for future work.


\bibliography{references}

\appendix
\section{Proof of Proposition \ref{prop:DPvsStandardLength}}\label{app-proof-prop:DPvsStandardLength}

\begin{proof}
Fix a true parameter value $\theta>0$. For $\tau>0$, the
direction-preferring and standard acceptance regions coincide. For
$\tau\leq 0$, they either coincide or differ by replacing a portion in
the right tail of the standard acceptance region with a portion farther
to the left. More precisely, let
\[
L_\tau
=
\mathcal A_{\DPp}(\tau;r,\alpha)
\setminus
\mathcal A_{\ns}(\tau;\alpha),
\qquad
R_\tau
=
\mathcal A_{\ns}(\tau;\alpha)
\setminus
\mathcal A_{\DPp}(\tau;r,\alpha).
\]
Every point in $L_\tau$ is smaller than every point in $R_\tau$.
Moreover, because both acceptance regions have probability $1-\alpha$
under $F_\tau$,
\[
\mP_\tau(Y\in L_\tau)
=
\mP_\tau(Y\in R_\tau).
\]

Since $\theta>\tau$, the monotone likelihood ratio property implies that
$f_\theta(y)/f_\tau(y)$ is nondecreasing in $y$. It therefore follows
that
\[
\mP_\theta(Y\in R_\tau)
\geq
\mP_\theta(Y\in L_\tau),
\]
and hence
\[
\mP_\theta\!\left\{
Y\in\mathcal A_{\ns}(\tau;\alpha)
\right\}
\geq
\mP_\theta\!\left\{
Y\in\mathcal A_{\DPp}(\tau;r,\alpha)
\right\}
\]
for every $\tau\leq0$. The same inequality holds with equality for
$\tau>0$. Therefore, Pratt's identity gives
\begin{align*}
\mE_{\theta}
\left[
\lambda\left\{
\mathcal A^{-1}_{\DPp}(Y;r,\alpha)
\right\}
\right]
&=
\int_{\mathbb R}
\mP_{\theta}
\left\{
Y\in\mathcal A_{\DPp}(\tau;r,\alpha)
\right\}
\,d\tau
\\
&\leq
\int_{\mathbb R}
\mP_{\theta}
\left\{
Y\in\mathcal A_{\ns}(\tau;\alpha)
\right\}
\,d\tau
\\
&=
\mE_{\theta}
\left[
\lambda\left\{
\mathcal I_{\mathrm{std}}(Y;\alpha)
\right\}
\right],
\end{align*}
where the last equality uses the fact that the exact inversion of the
standard acceptance regions is $\mathcal I_{\mathrm{std}}(Y;\alpha)$.
\end{proof}

\section{The Choice of the Inflation Parameter} \label{sect:choose_r}

\begin{figure}
 \centering
\includegraphics[width=0.4\paperwidth]{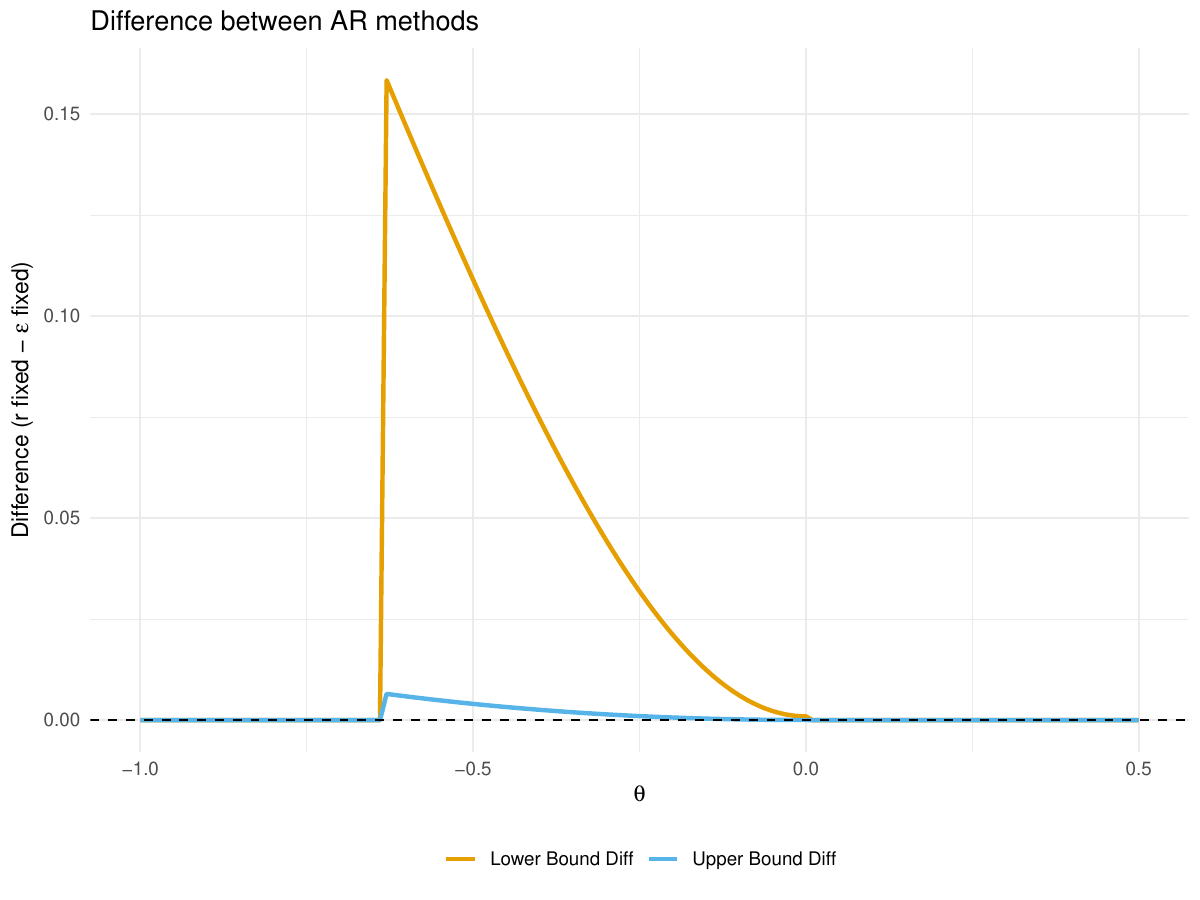}
 \caption{In the conditional setting, the difference between acceptance-region bounds obtained by fixing the inflation factor $r$ versus fixing the tail probability $\epsilon$. The curves show the differences in the lower and upper bounds as functions of $\theta$. Fixing $r$ results in a substantially higher lower bound compared to fixing $\epsilon$, whereas the corresponding upper bound is only slightly larger. \rev{The values $r=1.30$ and $\epsilon=1.09\times10^{-3}$ produce the same acceptance region at $\theta=0$ for $\alpha=0.05$ and $c=1.96$.}}
 \label{fig:diff_ar}
\end{figure}

\rev{For a given confidence level $1-\alpha$, the only free parameter in the direction-preferring construction is the inflation factor $r$. It controls the trade-off between the smallest observed $Y$ for which the preferred sign is determined and the CI length for observations in the non-preferred direction. The corresponding tail allocations are the derived quantities $\epsilon_{marg}(r,\alpha)$ and $\epsilon_{cond}(r,c,\alpha)$. In the marginal setting, fixing $r$ is equivalent to fixing $\epsilon_{marg}$, because the standard acceptance-region length is constant. In the conditional setting, fixing a tail allocation across $\theta$ can produce substantially more than the intended $r$-fold inflation away from zero, with only a minor improvement in one endpoint. We therefore fix $r$ throughout; see Figure~\ref{fig:diff_ar}.}

\begin{figure}
 \centering
 \includegraphics[width=1\textwidth]{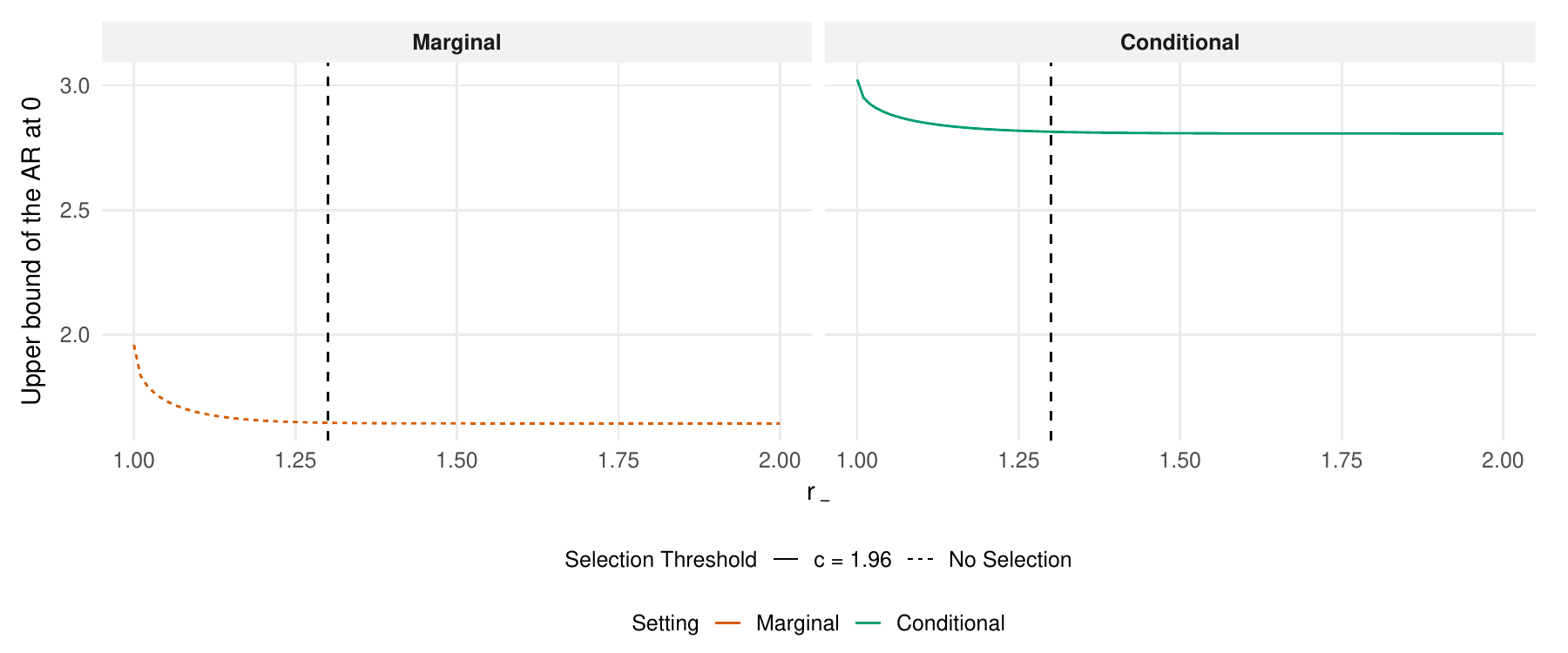}
 \caption{The minimal $y$ (assuming a normal distribution) for which the sign is determined for the \DPp CI as a function of $r$ in the marginal and conditional settings; in both settings $\alpha=0.05$. The truncation value used for the conditional setting is $c=1.96$. The vertical line is the \rev{recommended value $r=1.30$, which corresponds to $\epsilon_{marg}=2.82\times10^{-4}$ for the marginal CI and $\epsilon_{cond}=1.09\times10^{-3}$ for the conditional CI}.}
 \label{fig:choosing}
\end{figure}

\rev{We examine the choice in the marginal and conditional settings for $Y\sim N(\theta,1)$. In the conditional setting, the truncation value is $c=q_{0.975}=1.96$, a common two-sided selection threshold. Figure \ref{fig:choosing} shows that the gain from earlier sign determination diminishes rapidly beyond $r=1.30$. Marginally, $r=1.30$ yields strict preferred-sign determination for $y>1.65$, close to the limiting one-sided threshold $1.64$. Conditionally, with $c=1.96$, the thresholds are $3.02$ for $r=1$, $2.81$ for $r=1.30$, and $2.81$ for $r=1.50$, compared with the limiting threshold $2.81$. The corresponding allocations at $r=1.30$ are $\epsilon_{marg}=2.82\times10^{-4}$ and $\epsilon_{cond}=1.09\times10^{-3}$. We therefore recommend $r=1.30$. As a ratio of lengths, $r$ is dimensionless and unchanged by rescaling $Y$.}

\section{Complete Description of CIs} \label{sect:description_full}

\rev{Throughout this section, we use the method labels $\mathrm{std}$, $\mathrm{MP}$, $\DPp$, and $\mathrm{DP}^{-}$. We write $\Am_M(\theta;\cdot)$ and $\Imar_M(y;\cdot)$ for marginal acceptance regions and CIs, and $\Ac_M(\theta;\cdot)$ and $\Ic_M(y;\cdot)$ for their conditional counterparts. Arguments after the semicolon are ordered as $r$ (when applicable), $c$ (in the conditional setting), and $\alpha$. To agree with the main text, the compact DP CI notation retains $r$ in the subscript and suppresses the fixed conditional threshold $c$. Fixed values of $c$ and $\alpha$ are also suppressed in auxiliary endpoint notation.}

\subsection{Standard}

\subsubsection{Conditional Setting}

{Recall that $\mathcal S_c=(-\infty,-c)\cup(c,\infty)$, and let $|B|$ denote the Lebesgue measure of $B\subseteq\mathcal S_c$. Following the shortest-acceptance-region construction of \citet[Section~2]{weinstein2013selection}, the conditional standard acceptance region is the highest-density region
\begin{equation}\label{eq:acceptance_shortest}
\Ac_{\ns}(\theta;c,\alpha)=\{y\in\mathcal S_c:f^c_\theta(y)\geq\xi_\theta\},
\qquad
\mP_\theta\{Y\in\Ac_{\ns}(\theta;c,\alpha)\mid |Y|>c\}=1-\alpha,
\end{equation}
where $\xi_\theta$ is chosen to give the displayed probability. For a symmetric location family, $\Ac_{\ns}(\theta;c,\alpha)=-\Ac_{\ns}(-\theta;c,\alpha)$.

For the normal model, let
\[
Q_c(\theta)=\mP_\theta(|Y|>c)
=1-\Phi(c-\theta)+1-\Phi(c+\theta).
\]
For $\theta\geq0$, define
\[
\begin{aligned}
z_\theta&=\Phi^{-1}\{1-\alpha Q_c(\theta)/2\},\\
h_\theta&=\Phi^{-1}\{\Phi(c-\theta)+(1-\alpha)Q_c(\theta)\},\\
w_\theta&=\Phi^{-1}\{[1+(1-\alpha)Q_c(\theta)]/2\}.
\end{aligned}
\]
Then
\begin{equation}\label{eq:acceptance_shortest_normal}
\Ac_{\ns}(\theta;c,\alpha)=
\begin{cases}
(\theta-z_\theta,-c)\cup(c,\theta+z_\theta),
&0\leq\theta<\theta_1,\\
(c,\theta+h_\theta),
&\theta_1\leq\theta<\theta_2,\\
(\theta-w_\theta,\theta+w_\theta),
&\theta\geq\theta_2,
\end{cases}
\end{equation}
where $\theta_1$ and $\theta_2$ are the positive solutions of
\begin{align}
\Phi(c+\theta_1)-\Phi(c-\theta_1)
&=(1-\alpha)Q_c(\theta_1),\label{eq:shortest-theta1}\\
2\Phi(\theta_2-c)-1
&=(1-\alpha)Q_c(\theta_2).\label{eq:shortest-theta2}
\end{align}
The regions for $\theta<0$ are obtained by reflection.

The conditional standard CI is
\begin{equation}\label{eq:ci_shortest}
\Ic_{\ns}(y;c,\alpha)=\operatorname{conv}\{\theta:y\in\Ac_{\ns}(\theta;c,\alpha)\}.
\end{equation}
For the normal model and $y>c$, write
\[
\Ic_{\ns}(y;c,\alpha)
=[L_{\ns}^c(y),U_{\ns}^c(y)]
\]
and define
\[
x_1=c+2\theta_1,\qquad x_2=2\theta_2-c.
\]
The upper endpoint $U_{\ns}^c(y)$ is the solution of
\begin{equation}\label{eq:shortest-upper-root}
2\Phi\{U_{\ns}^c(y)-y\}-1
=(1-\alpha)Q_c\{U_{\ns}^c(y)\}.
\end{equation}
The lower endpoint is the solution, in the corresponding acceptance-region regime, of
\begin{equation}\label{eq:shortest-lower-root}
\begin{cases}
2[1-\Phi\{y-L_{\ns}^c(y)\}]
=\alpha Q_c\{L_{\ns}^c(y)\},
&c<y<x_1,\\
\Phi\{y-L_{\ns}^c(y)\}-\Phi\{c-L_{\ns}^c(y)\}
=(1-\alpha)Q_c\{L_{\ns}^c(y)\},
&x_1\leq y<x_2,\\
2\Phi\{y-L_{\ns}^c(y)\}-1
=(1-\alpha)Q_c\{L_{\ns}^c(y)\},
&y\geq x_2.
\end{cases}
\end{equation}
For $y<-c$, $\Ic_{\ns}(y;c,\alpha)=-\Ic_{\ns}(-y;c,\alpha)$. These expressions correspond to equations~(2)--(7) of \citet{weinstein2013selection}; their Appendix~A.1 provides the derivation.}

\subsection{Modified Pratt (MP)}

\subsubsection{Marginal Setting}\label{sect:marginal_ci_pratt}

{The modified Pratt construction bounds the acceptance-region length by $r$ times the length of the standard acceptance region while improving sign determination \citep{benjamini1998confidence}. Let $e=\epsilon_{marg}(r,\alpha)$. The marginal MP acceptance region is
\begin{equation}\label{eq:acceptance_mp}
\Am_{\mathrm{MP}}(\theta;r,\alpha)=
\begin{cases}
(\theta-q_{1-e},\theta+q_{1-\alpha+e}),&\theta<0,\\
(\theta-q_{1-\alpha/2},\theta+q_{1-\alpha/2}),&\theta=0,\\
(\theta-q_{1-\alpha+e},\theta+q_{1-e}),&\theta>0.
\end{cases}
\end{equation}
For $y\geq0$, its convex-hull inversion is
\begin{equation}\label{eq:ci_mp}
\Imar_{\mathrm{MP}}(y;r,\alpha)=
\begin{cases}
[y-q_{1-\alpha+e},y+q_{1-\alpha+e}],
&0\leq y<q_{1-\alpha+e},\\
[0,y+q_{1-\alpha+e}],
&q_{1-\alpha+e}\leq y<q_{1-\alpha/2},\\
(0,y+q_{1-\alpha+e}],
&q_{1-\alpha/2}\leq y<q_{1-e},\\
[y-q_{1-e},y+q_{1-\alpha+e}],
&y\geq q_{1-e},
\end{cases}
\end{equation}
and $\Imar_{\mathrm{MP}}(-y;r,\alpha)=-\Imar_{\mathrm{MP}}(y;r,\alpha)$. At $r=1$, $\epsilon_{marg}(1,\alpha)=\alpha/2$ and the MP CI coincides with the standard CI; the limiting choice $r=\infty$ gives the Pratt CI \citep{benjamini1998confidence,pratt1961length}.}

\subsubsection{Conditional Setting}

{The conditional MP construction of \citet[Section~3]{weinstein2013selection} is defined through its acceptance regions. Let
\[
\ell_{\ns}^c(\theta)=|\Ac_{\ns}(\theta;c,\alpha)|.
\]
At $\theta=0$, set $\Ac_{\mathrm{MP}}(0;r,c,\alpha)=\Ac_{\ns}(0;c,\alpha)$. For $\theta<0$, write
\[
\Ac_{\mathrm{MP}}(\theta;r,c,\alpha)
=\bigl(\ell_{\mathrm{MP}}^c(\theta;r),u_{\mathrm{MP}}^c(\theta;r)\bigr)\cap\mathcal S_c,
\]
where the endpoints solve
\begin{align}
F^c_\theta\{\ell_{\mathrm{MP}}^c(\theta;r)\}
+1-F^c_\theta\{u_{\mathrm{MP}}^c(\theta;r)\}&=\alpha,\\
\left|\bigl(\ell_{\mathrm{MP}}^c(\theta;r),u_{\mathrm{MP}}^c(\theta;r)\bigr)
\cap\mathcal S_c\right|&=r\ell_{\ns}^c(\theta),
\end{align}
with the solution extending farther into the negative tail. For $\theta>0$, define the region by reflection:
\begin{equation}\label{eq:acceptance_cmp}
\Ac_{\mathrm{MP}}(\theta;r,c,\alpha)=-\Ac_{\mathrm{MP}}(-\theta;r,c,\alpha).
\end{equation}
The conditional MP CI is generally defined as the convex hull of the exact inversion,
\begin{equation}\label{eq:ci_cmp}
\Ic_{\mathrm{MP}}(y;r,c,\alpha)=\operatorname{conv}\{\theta:y\in\Ac_{\mathrm{MP}}(\theta;r,c,\alpha)\}.
\end{equation}

For the normal model, the following gives the constructive form used in our implementation. Let $\widetilde\theta_1>0$ solve
\begin{equation}\label{eq:mp-theta-transition}
\Phi\{c+r\ell_{\ns}^c(\widetilde\theta_1)-\widetilde\theta_1\}
-\Phi(c-\widetilde\theta_1)
=(1-\alpha)Q_c(\widetilde\theta_1),
\end{equation}
and set
\[
z_c=\Phi^{-1}\{1-\alpha Q_c(0)/2\}.
\]
Let $\widetilde x_1<z_c<\widetilde x_2$ be the two solutions of
\begin{equation}\label{eq:mp-x-transitions}
1-\Phi(x)+1-\Phi\{2c+r\ell_{\ns}^c(0)-x\}
=\alpha Q_c(0),
\end{equation}
and define
\[
x_3=c+r\ell_{\ns}^c(\widetilde\theta_1).
\]
For compactness, define
\begin{align}
G_{\mathrm{MP}}(\theta;y)
&=1-\Phi(y-\theta)
+1-\Phi\{\theta-y+2c+r\ell_{\ns}^c(\theta)\}
-\alpha Q_c(\theta),\\
H_{\mathrm{MP}}(\theta;y)
&=\Phi(y-\theta)
-\Phi\{y-r\ell_{\ns}^c(\theta)-\theta\}
-(1-\alpha)Q_c(\theta),\\
K_{\mathrm{MP}}(\theta;y)
&=\Phi\{y+r\ell_{\ns}^c(\theta)-\theta\}
-\Phi(y-\theta)
-(1-\alpha)Q_c(\theta).
\end{align}
Let $L_{\mathrm{MP},-}^c(y;r)$ be the largest root of $G_{\mathrm{MP}}(\theta;y)=0$ on $[-\widetilde\theta_1,0]$, let $L_{\mathrm{MP},+}^c(y;r)$ be its smallest root on $[0,\widetilde\theta_1]$, let $L_{\mathrm{MP},t}^c(y;r)$ be the smallest root of $H_{\mathrm{MP}}(\theta;y)=0$ on $[\widetilde\theta_1,\infty)$, and let $U_{\mathrm{MP}}^c(y;r)$ be the largest positive root of $K_{\mathrm{MP}}(\theta;y)=0$. Then, for $y>c$ and ordered transition values,
\begin{equation}\label{eq:ci_cmp_normal}
\Ic_{\mathrm{MP}}(y;r,c,\alpha)=
\begin{cases}
[L_{\mathrm{MP},-}^c(y;r),U_{\mathrm{MP}}^c(y;r)],
&c<y<\widetilde x_1,\\
[0,U_{\mathrm{MP}}^c(y;r)],
&\widetilde x_1\leq y<z_c,\\
(0,U_{\mathrm{MP}}^c(y;r)],
&z_c\leq y<\widetilde x_2,\\
[L_{\mathrm{MP},+}^c(y;r),U_{\mathrm{MP}}^c(y;r)],
&\widetilde x_2\leq y<x_3,\\
[L_{\mathrm{MP},t}^c(y;r),U_{\mathrm{MP}}^c(y;r)],
&y\geq x_3.
\end{cases}
\end{equation}
For $y<-c$, $\Ic_{\mathrm{MP}}(y;r,c,\alpha)=-\Ic_{\mathrm{MP}}(-y;r,c,\alpha)$. If the transition values are not ordered, equation~\eqref{eq:ci_cmp} remains the definition and is evaluated by direct inversion. At $r=1$, the construction reduces to the conditional standard CI.}

\subsection{Direction-Preferring CIs}

{The direction-preferring construction adapts the MP construction by using the inflated acceptance region only in the nonpreferred parameter direction and reverting to the standard acceptance region once the sign is determined by both constructions. The marginal DP$^+$ CI $\mathcal I_{\DPp,r}(y;\alpha)$ is given explicitly in \eqref{eq:pp-marginal}. The corresponding conditional acceptance region $\Ac_{\DPp}(\theta;r,c,\alpha)$ is given in \eqref{eq:spcmp_acceptance}.}

\subsubsection{Conditional Setting}\label{sect:pp_ci}

{The conditional DP$^+$ CI is generally defined by
\begin{equation}\label{eq:spcmp_CI}
\mathcal I^c_{\DPp,r}(y;\alpha)
=\operatorname{conv}\{\theta:y\in\Ac_{\DPp}(\theta;r,c,\alpha)\}.
\end{equation}
Let
\[
a_{\DPp,-}^c(\theta;r)=\inf\Ac_{\DPp}(\theta;r,c,\alpha),
\qquad
a_{\DPp,+}^c(\theta;r)=\sup\Ac_{\DPp}(\theta;r,c,\alpha),
\]
and similarly let $a_{\ns,-}^c(\theta)$ and $a_{\ns,+}^c(\theta)$ denote the outer boundaries of $\Ac_{\ns}(\theta;c,\alpha)$. Write
\[
\theta_r=\widetilde\theta^-_1(r).
\]
Let $L_{\DPp}^c(y;r)$ be the smallest root of
\[
a_{\DPp,+}^c(\theta;r)=y
\]
on $[\theta_r,0]$, and let $U_{\DPp}^c(y;r)$ be the largest root of
\[
a_{\DPp,-}^c(\theta;r)=y
\]
on that interval. When
\[
a_{\DPp,-}^c(\theta_r;r)
<a_{\DPp,-}^c(0;r)
<a_{\ns,-}^c(0)
<-c<c
<a_{\DPp,+}^c(0;r)
<a_{\ns,+}^c(0),
\]
as in the numerical settings considered here, the package implements the following seven-regime inversion:
\begin{equation}\label{eq:spcmp_CI_normal}
\mathcal I^c_{\DPp,r}(y;\alpha)=
\begin{cases}
[L_{\ns}^c(y),U_{\ns}^c(y)],
&y<a_{\DPp,-}^c(\theta_r;r),\\
[L_{\ns}^c(y),U_{\DPp}^c(y;r)],
&a_{\DPp,-}^c(\theta_r;r)\leq y<a_{\DPp,-}^c(0;r),\\
[L_{\ns}^c(y),0],
&a_{\DPp,-}^c(0;r)\leq y<a_{\ns,-}^c(0),\\
[L_{\ns}^c(y),U_{\ns}^c(y)],
&a_{\ns,-}^c(0)\leq y<-c,\\
[L_{\DPp}^c(y;r),U_{\ns}^c(y)],
&c<y\leq a_{\DPp,+}^c(0;r),\\
(0,U_{\ns}^c(y)],
&a_{\DPp,+}^c(0;r)<y\leq a_{\ns,+}^c(0),\\
[L_{\ns}^c(y),U_{\ns}^c(y)],
&a_{\ns,+}^c(0)<y.
\end{cases}
\end{equation}
If these transition values are not ordered, equation~\eqref{eq:spcmp_CI} remains the definition and is evaluated by direct inversion of \eqref{eq:spcmp_acceptance}. The DP$-$ CI is obtained by reflection:
\[
\mathcal I^c_{\mathrm{DP}^{-},r}(y;\alpha)
=-\mathcal I^c_{\DPp,r}(-y;\alpha).
\]}


\section{Sign Determination Comparison}

The cumulative conditional distribution of \rev{$Y\mid |Y|>c$} can be written in terms of the non-conditional distribution $F_\theta$,

$$\Fc (y) = \begin{cases}
  F_\theta \left( y \right) / t_\theta(c)  & \text{if}   \; y < -c  \\
  \rev{\left(F_\theta(y)+t_\theta(c)-1\right)/t_\theta(c)} & \text{if} \quad y>c.
\end{cases} $$

where $P(|Y| > c) = t_\theta(c) = 1 - F_\theta(c) + F_\theta (-c)$, the selection probability according to a selection threshold of $c$.

By inverting it, we obtain the quantile function,

\begin{equation} \label{eq:cond_quantile}
 \left( \Fc\right)^{-1} (p) = \begin{cases}
  F_\theta^{-1} \left( p \cdot t_\theta(c)\right) & \text{if} \;  p \leq \Fc(-c)  \\
  F_\theta^{-1} \left( p \cdot t_\theta(c) + 1 -t_\theta(c) \right) & \text{if} \;    p > \Fc(c).
\end{cases}
\end{equation}

\subsection{Proof of Theorem \ref{thm:SD-Conditional-BY}}\label{app-proof-thm:SD-Conditional-BY}
\rev{By construction, the upper bound at $\theta=0$ is the largest upper bound among the direction-preferring acceptance regions for $\theta\leq0$. Hence} the positive sign is determined for any $Y$ that is above the upper bound on the acceptance region at $\theta=0$. Therefore, we need to show that the upper bound at $\theta=0$ for the conditional acceptance region at level $\alpha$ is at least as large as that of the marginal acceptance region at level \rev{$\alpha\widehat{\pi}_m$}.

To see this, note that the upper bound of the marginal acceptance region at $\theta=0$, \rev{$\Am_{\DPp}(0;r,\widetilde\alpha)$}, is monotone decreasing in $\widetilde\alpha$. Given that $\widehat{\pi}_m\geq t_0$, it is enough to show that the upper bound at $\theta=0$ for the conditional acceptance region at level $\alpha$ is at least as large as that of \rev{$\Am_{\DPp}(0;r,\alpha t_0)$}.

The length of the marginal standard (shortest) acceptance region at level $\alpha t_0$ is $2q_{1-\alpha t_0/2}$.
The upper bound of \rev{$\Am_{\DPp}(0;r,\alpha t_0)$} is $q_{1-\alpha t_0+\epsilon}$, where \rev{$\epsilon=\epsilon_{marg}(r,\alpha t_0)$} is the solution to
\[
\rev{q_{1-\alpha t_0+\epsilon}-q_\epsilon=2r q_{1-\alpha t_0/2}}.
\]

The length of the \rev{level-$\alpha$ conditional} standard acceptance region is $2q_{1-\alpha t_0/2}-2c$, using \eqref{eq:cond_quantile}.
For the level $\alpha$ conditional acceptance region at $\theta=0$, again using \eqref{eq:cond_quantile}, the upper bound is $q_{1-\alpha t_0+\widetilde\epsilon}$, where \rev{$\widetilde\epsilon=t_0\epsilon_{cond}(r,c,\alpha)$} is the solution to
\[
\rev{q_{1-\alpha t_0+\widetilde\epsilon}-q_{\widetilde\epsilon}-2c
=2r\bigl(q_{1-\alpha t_0/2}-c\bigr)},
\]
i.e.,
\[
\rev{q_{1-\alpha t_0+\widetilde\epsilon}-q_{\widetilde\epsilon}
=2r q_{1-\alpha t_0/2}-2(r-1)c}.
\]

{To compare the two allocations, define $G(\eta)=q_{1-\alpha t_0+\eta}-q_\eta$ for $0\leq\eta\leq\alpha t_0/2$. By symmetry,
\[
G(\eta)=q_{1-\alpha t_0+\eta}+q_{1-\eta},
\]
and
\[
G'(\eta)=\frac{1}{f(q_{1-\alpha t_0+\eta})}
-\frac{1}{f(q_{1-\eta})}\leq0,
\]
because $q_{1-\alpha t_0+\eta}\leq q_{1-\eta}$ and $f$ is nonincreasing on $[0,\infty)$. Thus $G$ is nonincreasing. If $r>1$, then $G(\widetilde\epsilon)<G(\epsilon)$, so $\widetilde\epsilon>\epsilon$. If $r=1$, both constructions reduce to the corresponding standard acceptance region and $\widetilde\epsilon=\epsilon=\alpha t_0/2$. Hence $\widetilde\epsilon\geq\epsilon$ for every $r\geq1$.} Therefore, the upper bound at $\theta=0$ for the conditional acceptance region at level $\alpha$ is at least as large as that of \rev{$\Am_{\DPp}(0;r,\alpha t_0)$}.

\subsection{Proof of Corollary \ref{cor:SD-Conditional-BY}}\label{app-proof-cor:SD-Conditional-BY}

{Let $\widehat{\pi}_m=|S(\boldsymbol{Y})|/m$ and define
\[
A_m=\{0\notin\mathcal I^c_{\DPp,r}(Y_i;\alpha),\ Y_i>0\},\quad
B_m=\{0\notin\mathcal I_{\DPp,r}(Y_i;\alpha\widehat{\pi}_m),\ Y_i>0\}.
\]
Let $b_c$ be the upper endpoint at $\theta=0$ of the conditional direction-preferring acceptance region. For an error level $\gamma$, define the upper endpoint of the marginal direction-preferring acceptance region at $\theta=0$ by
\[
b(\gamma):=\sup\Am_{\DPp}(0;r,\gamma)
=q_{1-\gamma+\epsilon_{marg}(r,\gamma)}.
\]
Thus the random BY-adjusted endpoint is $b_m:=b(\alpha\widehat{\pi}_m)$, and its limiting value is $b_S:=b(\alpha\pi_S)$. Since $\pi_S\geq t_0$, the endpoint comparison in the proof of Theorem \ref{thm:SD-Conditional-BY}, applied with the constant $\pi_S$ in place of $\widehat{\pi}_m$, gives $b_S\leq b_c$. By the assumed continuity and the continuous mapping theorem,
\[
b_m\xrightarrow{p}b_S.
\]
For positive observations, apart from boundary events of probability zero, $A_m=\{Y_i>b_c\}$ and $B_m=\{Y_i>b_m\}$. If $b_S<b_c$, it follows immediately that $\mP(A_m\cap B_m^c)\to0$. If $b_S=b_c$, then for every $\varepsilon>0$,
\[
\mP(A_m\cap B_m^c)
\leq \mP\{|b_m-b_c|>\varepsilon\}
   +\mP\{b_c<Y_i\leq b_c+\varepsilon\}.
\]
The first term tends to zero, and the second tends to zero as $\varepsilon\downarrow0$ because $Y_i$ has a continuous distribution. Thus $\mP(A_m\cap B_m^c)\to0$, so
\[
\mP(B_m)\geq\mP(A_m)-o(1).
\]
Taking lower limits proves the result.}

\section{Additional FCR vs Conditional Simulations}
\label{sect:additional_fcr_sims}

This appendix reports additional simulation results extending the main experiments
to different problem dimensions.
Specifically, we consider settings with $p=20$ and $p=50$ parameters,
while keeping the remaining aspects of the data-generating process unchanged.

Data are generated as $Y_i \sim N(\theta_i,1)$ with AR(1) dependence
$\Sigma_{ij} = 0.7^{|i-j|}$.
\rev{The proportion of non-zero parameters varies over $0.05$, $0.10$, $0.20$, and $0.50$, and all non-zero effects share a common value $\theta\in\{-3.00,-1.50,-0.50,0.00,0.50,1.50,3.00\}$. The figures use AR(1) correlation $0.70$, and each configuration uses 1000 simulation replicates.}
Selection is performed using the same threshold $|Y_i| > 1.96$ as in the main text.
We compare conditional CIs and BY-adjusted marginal CIs
constructed using the MP, \DPp, and standard procedures.

\begin{figure}[htbp]
\centering
\includegraphics[width=0.9\textwidth]{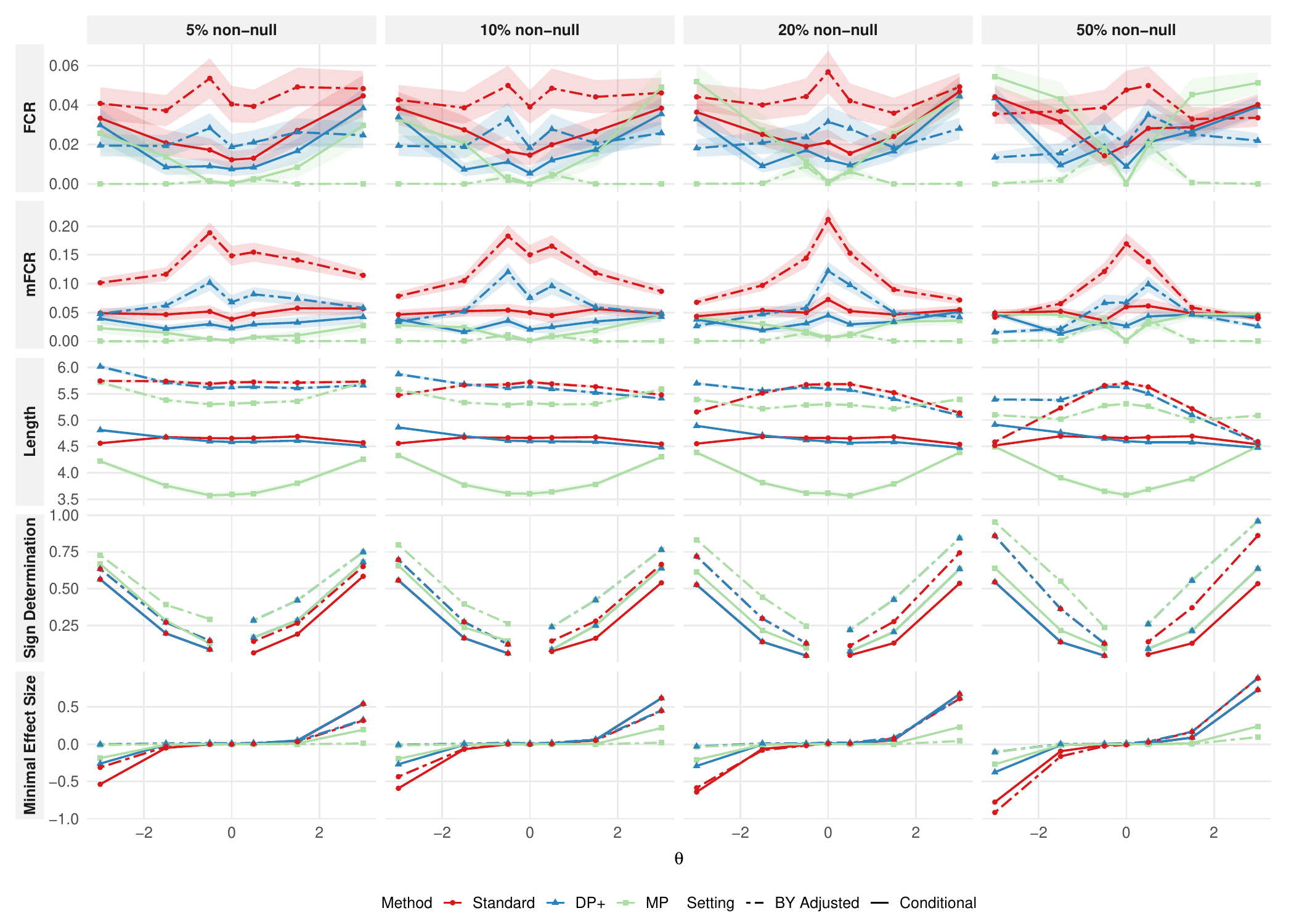}
\caption{
Comparison of conditional and BY-adjusted CIs for $p=20$ parameters.
Data are generated as $Y_i \sim N(\theta_i,1)$ with AR(1) dependence
$Cor(Y_i,Y_j)=0.7^{|i-j|}$.
Columns correspond to the proportion of non-zero parameters,
and the x-axis shows the common non-zero effect size $\theta$.
Rows display, from top to bottom, FCR, mFCR, interval length, the proportion of sign-determined parameters, and minimal effect size. Conditional CIs are typically shorter and control the mFCR, whereas BY-adjusted CIs control the FCR and generally yield more sign determinations but may inflate the mFCR in sparse regimes.
}
\label{fig:appendix_m20}
\end{figure}

\begin{figure}[htbp]
\centering
\includegraphics[width=0.9\textwidth]{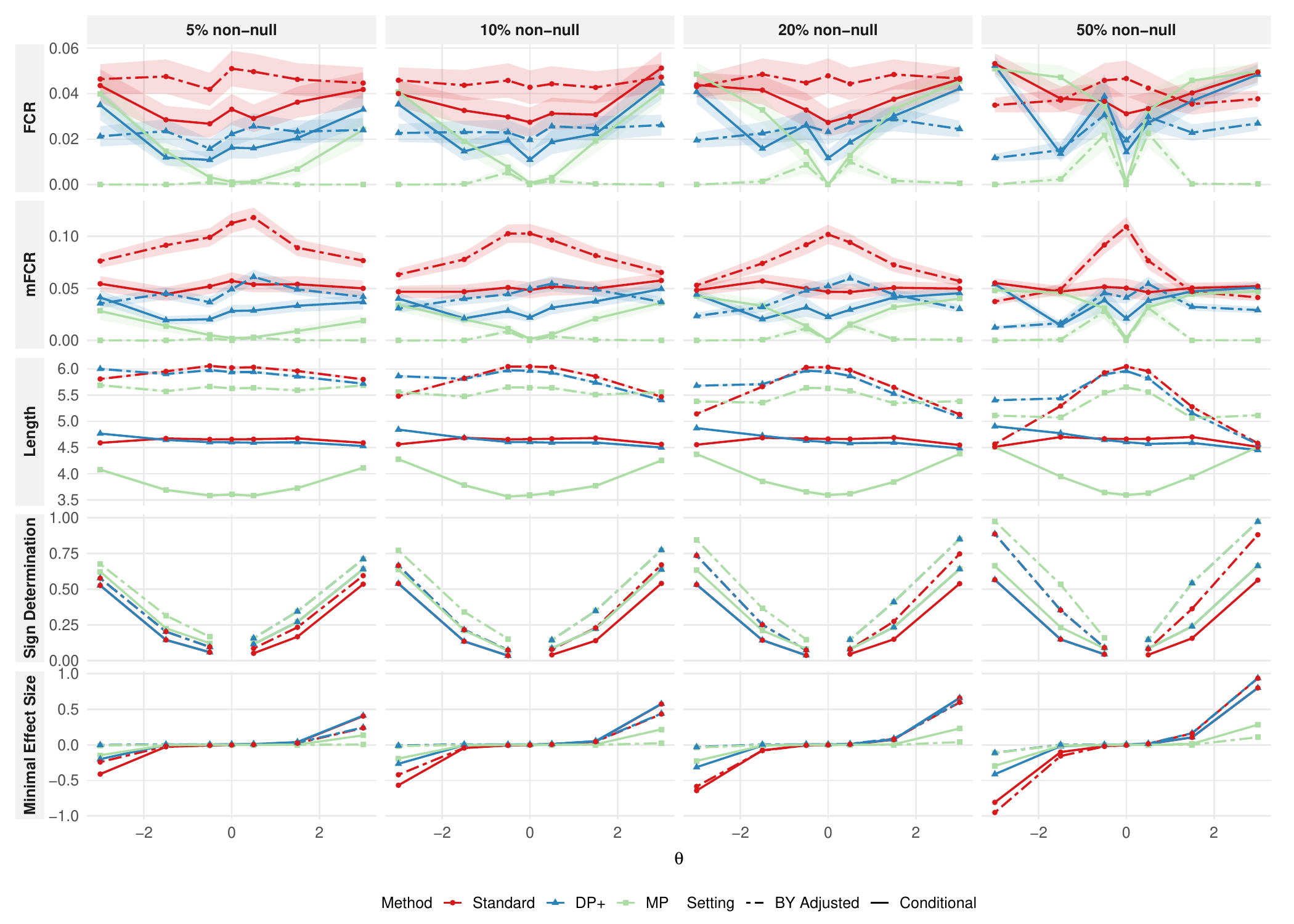}
\caption{
Comparison of conditional and BY-adjusted CIs for $p=50$ parameters.
The simulation setup matches that of Figure~\ref{fig:appendix_m20},
with AR(1) dependence and a common non-zero effect size $\theta$.
Increasing the dimension does not qualitatively change the observed patterns:
conditional CIs remain shorter and control the mFCR,
whereas BY-adjusted CIs achieve higher sign determination at the cost of mFCR inflation
in sparse settings.
}
\label{fig:appendix_m50}
\end{figure}

\Cref{fig:appendix_m20} and \Cref{fig:appendix_m50} summarize the results.
The qualitative patterns observed in the main text persist across both dimensions.
In particular, conditional CIs remain shorter than their BY-adjusted counterparts
across most configurations, while controlling the mFCR uniformly.
In contrast, BY-adjusted intervals continue to exhibit inflated mFCR in sparse regimes,
with the inflation diminishing as the signal strength or proportion of non-null parameters increases.

Similarly, BY-adjusted CIs achieve a higher proportion of sign-determined parameters, \rev{although the selection rule, and hence the selected set, is the same for both approaches. The difference is due to the CI endpoints rather than more frequent selection.}
Conditional CIs yield fewer sign determinations,
but those determinations are associated with larger minimal effect sizes.
Overall, increasing the dimension from $p=20$ to $p=50$ does not materially alter the relative behavior
of the two approaches, indicating that the phenomena observed in the main text are not artifacts
of a particular problem size.

\section{One-Sided Conditional CIs} \label{sect:conditional_os}

\begin{figure}
 \centering
 \includegraphics[width=\textwidth]{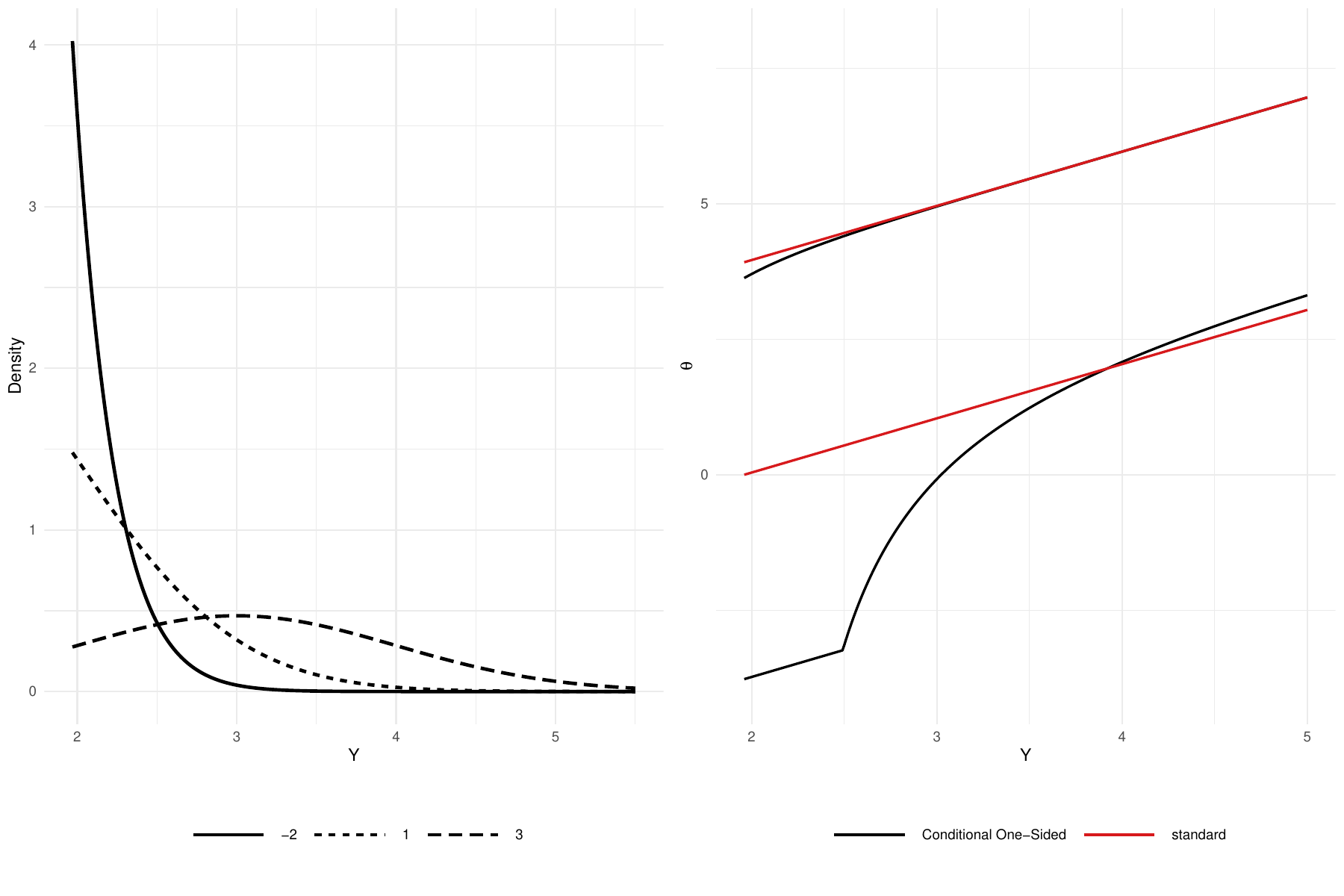}
 \caption{The density plots and CI of one-sided conditional CIs, $c=1.96$, $\alpha=0.05$. The one-sided conditional CIs are \rev{extremely} wide when the sign is not determined; this is due to all of the conditional probability mass being concentrated above $c$, regardless \rev{of} the value of $\theta$.}\label{fig:os_ci_and_dens}
\end{figure}

\rev{An alternative to constructing direction-preferring conditional CIs is to distribute the error probability between one-sided conditional bounds according to the analyst's preferences. For selection $Y>c$, define
\begin{equation}\label{eq:acceptance_os_cond}
F^{c,+}_\theta(y)
=\mP_\theta(Y\leq y\mid Y>c)
=\frac{F(y-\theta)-F(c-\theta)}{1-F(c-\theta)},\qquad y>c.
\end{equation}
The equal-tailed conditional CI used in the accompanying implementation is obtained by inversion:
\begin{equation}\label{eq:ci_os_cond}
\Ic_{\mathrm{OS}^{+}}(y;c,\alpha)
=[L_{\mathrm{OS}^{+}}^c(y),U_{\mathrm{OS}^{+}}^c(y)],
\end{equation}
where $L_{\mathrm{OS}^{+}}^c(y)$ and $U_{\mathrm{OS}^{+}}^c(y)$ are the unique roots of
\[
F^{c,+}_{L_{\mathrm{OS}^{+}}^c(y)}(y)=1-\alpha/2,
\qquad
F^{c,+}_{U_{\mathrm{OS}^{+}}^c(y)}(y)=\alpha/2.
\]
Unequal tail allocations can be obtained by replacing $\alpha/2$ in the two equations by nonnegative allocations that sum to $\alpha$. Even without an extreme allocation, these CIs can be exceedingly long when the sign is not determined. This behavior is consistent with \cite{kivaranovic2021length}, who show that truncation can lead to infinite expected interval length. Figure \ref{fig:os_ci_and_dens} illustrates the phenomenon.}

\section{Additional SURPASS details}\label{app-sect:real_example}

\begin{figure}
 \centering
\includegraphics[width=\textwidth]{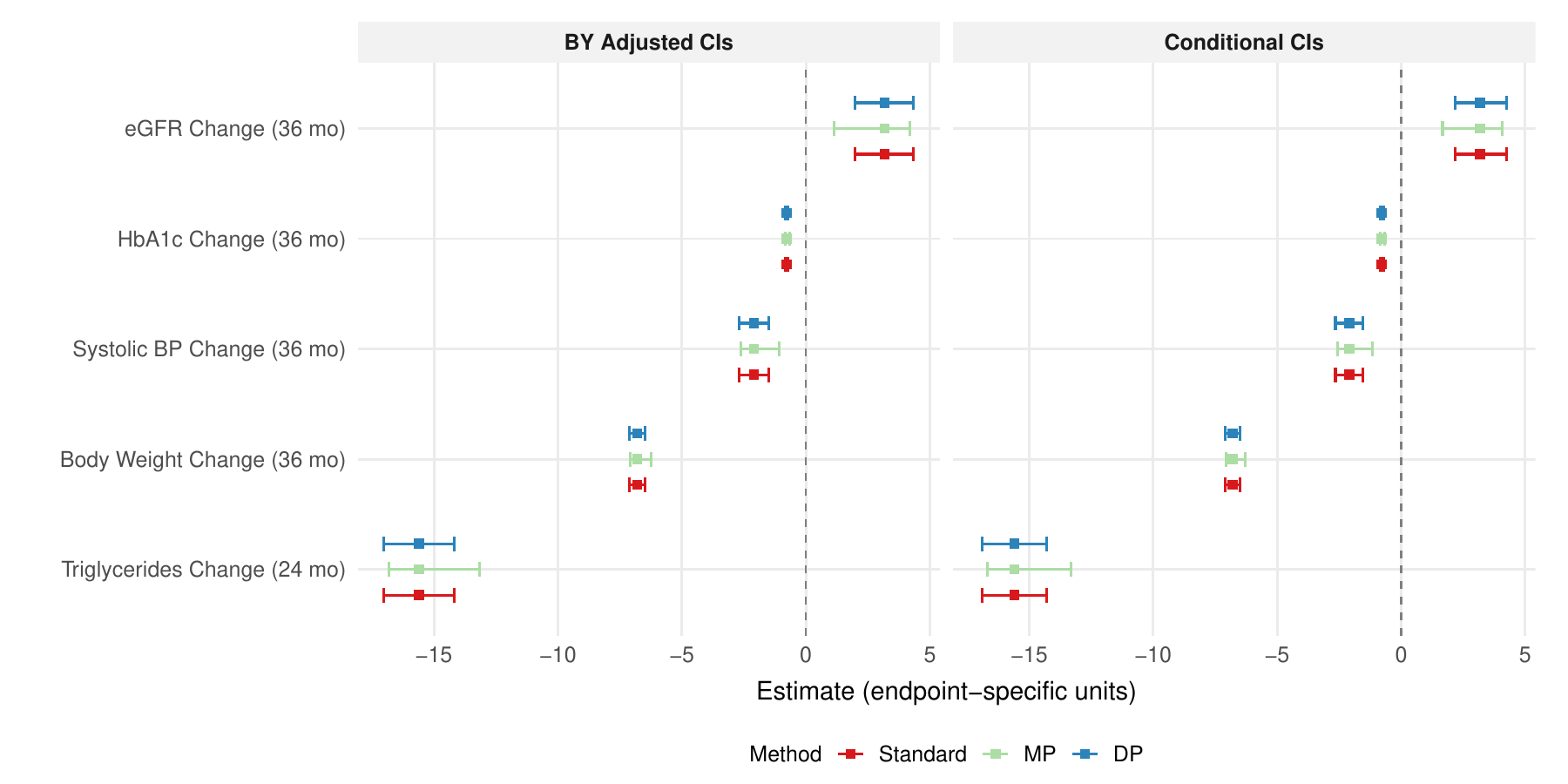}
 \caption{Comparing the DP, Standard and MP CIs for the five selected difference endpoints from the SURPASS-CVOT trial \rev{\cite{Nicholls2025Cardiovascular}}. The joint family across ratio and difference outcomes has $m=12$ secondary endpoints with $R=8$ selected, giving a BY-adjustment level of $\alpha_{\text{BY}}=(R/m)\alpha=8/12\times0.05\approx\rev{0.0333}$. The selection threshold is \rev{$c=q_{1-\alpha/2}=1.96$}; all CIs were constructed at confidence level $0.95$ and, when applicable, inflation parameter \rev{$r=1.30$}. Direction-preferring CIs use each endpoint's preferred direction (DP$-$ for HbA1c, body weight, triglycerides, and systolic BP; DP$^+$ for eGFR). Conditional CIs depend only on the selection threshold $c$ and not on $m$ or $R$. }
 \label{fig:surpass_diff_example}
\end{figure}

We provide here the results for the additional mean differences in the SURPASS trial. As seen for \rev{five of the six secondary mean-difference endpoints}, the differences are very large, leading to very narrow CIs as seen in \cref{fig:surpass_diff_example}.

\section{Identifying genotypes associated with sudden cardiac arrest}\label{sec-GWAS-example}

The objective in GWAS is to identify single nucleotide polymorphisms (SNPs) that show a statistical association with a specific trait or phenotype. This association is assessed by measuring the impact of the minor allele, in contrast to the major allele. In an effort to understand the genetic underpinnings of sudden cardiac arrest, particularly among patients with coronary artery disease, \cite{aouizerat2011gwas} conducted a GWAS contrasting 89 such patients with 520 healthy controls. The findings presented a significant association of 14 SNPs with sudden cardiac arrest. In total, $319,222$ SNPs were analyzed, with the $14$ SNPs being significant after a Bonferroni correction. They present odds ratios (ORs) along with $0.95$ unadjusted standard CIs, which substantially understate uncertainty after selection given that a selection occurred. For illustration, it is assumed that the analysts are primarily focused on SNPs associated with increased risk. 
\rev{ To} address this focus, we employ direction-preferring CIs, controlled for their multiplicity. Furthermore, we demonstrate two methods for adjusting for selection: \rev{conditional DP$^+$ CIs and BY-adjusted DP$^+$ CIs}.

\begin{figure}
 \centering
\includegraphics[width=\textwidth]{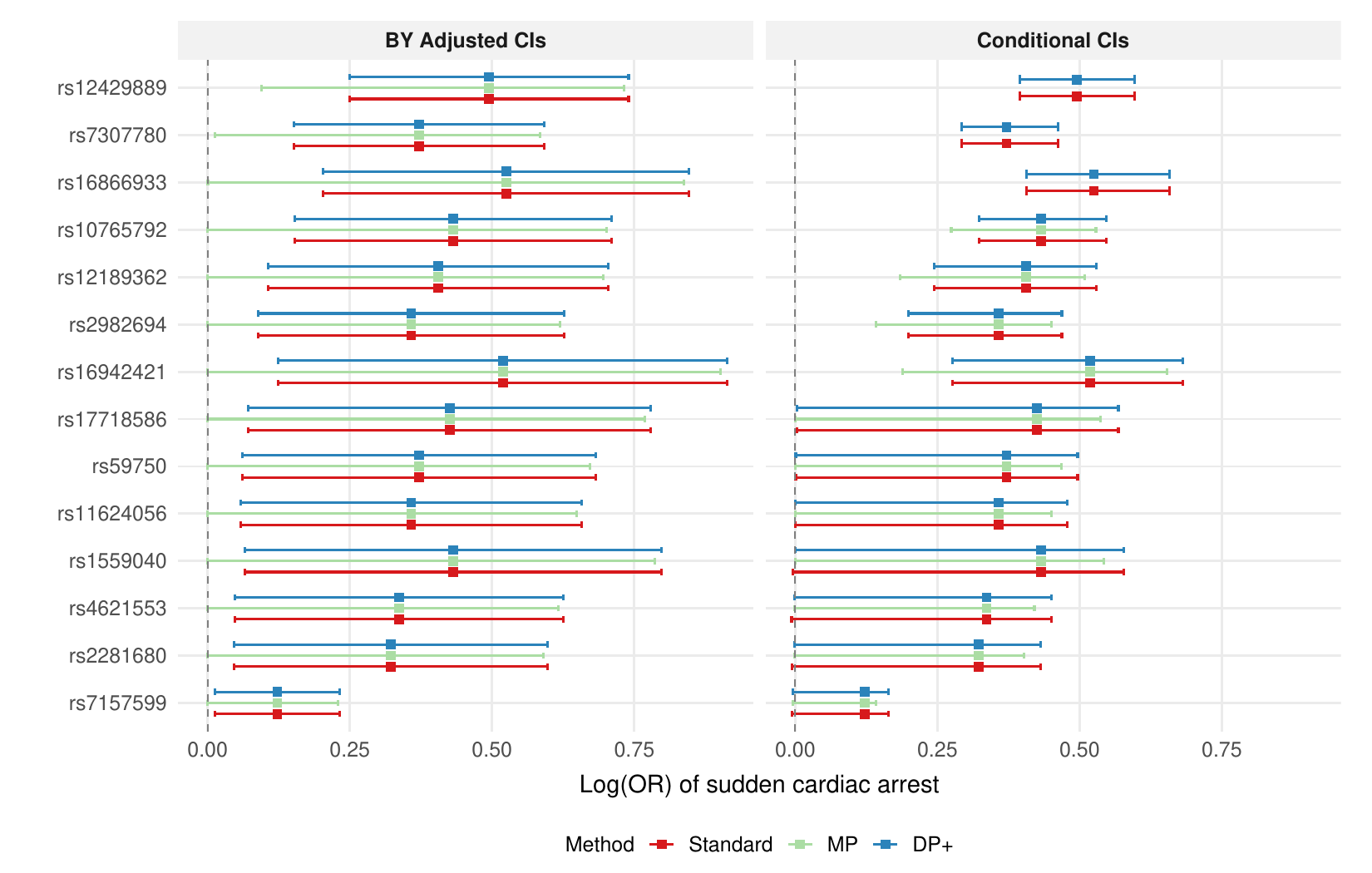}
 \caption{Comparing the DP, standard and MP CIs for the 14 selected SNPs from \cite{aouizerat2011gwas}. The selection threshold is \rev{$c=q_{1-0.05/319222}=5.11$}; all CIs were constructed at confidence level $0.95$ and, when applicable, inflation parameter \rev{$r=1.30$}. }
 \label{fig:gwas_example}
\end{figure}

The results are summarized in Table \ref{tab:gwas_example} and the CIs are depicted in Figure \ref{fig:gwas_example}. The conditional CIs are shorter on average and have higher average lower bounds than their BY-adjusted counterparts. Under weak sign determination, all three BY-adjusted methods determine the sign for all 14 SNPs. Among the conditional methods, the standard CI determines the sign for 9 SNPs, whereas the DP+ and MP CIs determine the sign for the same 11 SNPs. DP$^+$ provides stronger separation from zero than MP: its average lower bound is 0.152 versus 0.123 conditionally, and 0.103 versus 0.008 after BY adjustment. Under strict sign determination, however, the corresponding counts are 9 for conditional DP$^+$, 7 for conditional MP, and only 3 for BY-adjusted MP.

\begin{table}[]
\centering
\caption{Summary of the CI comparison for the
\cite{aouizerat2011gwas} example. Sign determination is weak, meaning
that an interval may have zero as an endpoint. The mean length and mean
lower bound are averages across all 14 CIs on the log(OR) scale.}
\begin{tabular}{ccccc}
\hline
Type &
Method &
\begin{tabular}[c]{@{}c@{}} Sign Determination\end{tabular} &
Mean Length &
Mean Lower Bound \\
\hline
\multirow{3}{*}{Conditional}
  & Standard          & 9  & 0.357 & 0.151 \\
  & DP$^+$ ($r=1.30$) & 11 & 0.355 & 0.152 \\
  & MP ($r=1.30$)    & 11 & 0.360 & 0.123 \\
\hline
\multirow{3}{*}{BY-adjusted}
  & Standard          & 14 & 0.576 & 0.103 \\
  & DP$^+$ ($r=1.30$) & 14 & 0.576 & 0.103 \\
  & MP ($r=1.30$)    & 14 & 0.662 & 0.008 \\
\hline
Unadjusted
  & Standard          & 14 & 0.238 & 0.271 \\
\hline
\end{tabular}
\label{tab:gwas_example}
\end{table}

\end{document}